\documentclass[prl,reprint,aps,superscriptaddress,nobibnotes,nofootinbib]{revtex4-2}

\usepackage{dcolumn}% Align table columns on decimal point
\usepackage{bm}% bold math
\usepackage[tight]{units}
\usepackage[colorlinks,citecolor=blue,urlcolor=blue,linkcolor=blue]{hyperref}
\usepackage{hyperref}% add hypertext capabilities+
\usepackage[normalem]{ulem}
\usepackage{verbatim}
\usepackage[dvipsnames]{xcolor}
\usepackage[tableposition=top,font=small,labelfont=bf]{caption}
\usepackage{graphicx, color}
\graphicspath{{fig/}}
\usepackage{subcaption}

\usepackage{amsmath}
\usepackage{wasysym}
\usepackage{xurl}
\usepackage{orcidlink}

\begin{document}

%\preprint{APS/123-QED}

\title{A Plastic Scintillation Muon Veto for Sub-Kelvin Temperatures}% 

\vspace{20em}

\newcommand{\addra}{\affiliation{Physik-Department, Technische Universit\"at M\"unchen, D-85748 Garching, Germany}}
\newcommand{\addrb}{\affiliation{IRFU, CEA, Universit\'{e} Paris-Saclay, F-91191 Gif-sur-Yvette, France}}
\newcommand{\addrc}{\affiliation{Atominstitut, Technische Universit\"at Wien, A-1020 Wien, Austria}}
\newcommand{\addrd}{\affiliation{Max-Planck-Institut f\"ur Physik, D-80805 M\"unchen, Germany}}
\newcommand{\addre}{\affiliation{LIBPhys-UC, Departamento de Fisica, Universidade de Coimbra, P-3004\,516 Coimbra, Portugal}}
\newcommand{\addrf}{\affiliation{Istituto Nazionale di Fisica Nucleare -- Sezione di Roma, I-00185 Roma, Italy}}
\newcommand{\addrg}{\affiliation{Istituto Nazionale di Fisica Nucleare -- Sezione di Roma "Tor Vergata", I-00133 Roma, Italy}}
\newcommand{\addrh}{\affiliation{Dipartimento di Fisica, Universit\`{a} di Roma "Tor Vergata", I-00133 Roma, Italy}}
\newcommand{\addri}{\affiliation{Consiglio Nazionale delle Ricerche, Istituto di Nanotecnologia, I-00185 Roma, Italy}}
\newcommand{\addrj}{\affiliation{Dipartimento di Fisica, Sapienza Universit\`{a} di Roma, I-00185 Roma, Italy}}
\newcommand{\addrk}{\affiliation{Institut f\"ur Hochenergiephysik der \"Osterreichischen Akademie der Wissenschaften, A-1050 Wien, Austria}}
\newcommand{\addrl}{\affiliation{IRFU (DPhP \& APC), Universit\'{e} Paris-Saclay, F-91191 Gif-sur-Yvette, France}}
\newcommand{\addrm}{\affiliation{Istituto Nazionale di Fisica Nucleare -- Sezione di Ferrara, I-44122 Ferrara, Italy}}
\newcommand{\addrn}{\affiliation{Istituto Nazionale di Fisica Nucleare -- Laboratori Nazionali del Gran Sasso, I-67100 Assergi (L’Aquila), Italy}}
\newcommand{\addro}{\affiliation{Dipartimento di Fisica, Universit\`{a} di Ferrara, I-44122 Ferrara, Italy}}

\addra
\addrb
\addrc
\addrd
\addre
\addrf
\addrg
\addrh
\addri
\addrj
\addrk
\addrl
\addrm
\addrn
\addro

\author{A.~Erhart\orcidlink{0000-0002-8721-177X}}
    \email[Corresponding author: ]{andreas.erhart@tum.de}
    \addra

\author{V.~Wagner\orcidlink{0000-0003-1845-4951}}
    \addra

\author{A.~Wex\orcidlink{0009-0003-5371-2466}}
    \addra

\author{C.~Goupy\orcidlink{0000-0003-4954-5311}}
    \addrb

\author{D.~Lhuillier\orcidlink{0000-0003-2324-0149}}
    \addrb

\author{E.~Namuth\orcidlink{0009-0001-9516-2058}}
    \addra

\author{C.~Nones}
    \addrb

\author{R.~Rogly}
    \addrb

\author{V.~Savu}
    \addrb

\author{M.~Schwarz\orcidlink{0000-0002-8360-666X}}
    \addra

\author{R.~Strauss}
    \addra

\author{M.~Vivier}
    \addrb

\author{H.~Abele}
\addrc 
        
\author{G.~Angloher}
\addrd 
        
\author{A.~Bento}
\addrd
\addre 
        
\author{J.~Burkhart\orcidlink{0000-0002-1989-7845}}
\addrk
        
\author{L.~Canonica}
\addrd
        
\author{F.~Cappella\orcidlink{0000-0003-0900-6794}}
\addrf 
        
\author{N.~Casali}
\addrf
        
\author{R.~Cerulli\orcidlink{0000-0003-2051-3471}}
\addrg
\addrh
        
\author{A.~Cruciani}
\addrf
        
\author{G.~del Castello\orcidlink{0000-0001-7182-358X}}
\addrf
\addrj 
        
\author{M.~del~Gallo~Roccagiovine}
\addrf
\addrj 
        
\author{A.~Doblhammer}
\addrc
        
\author{S.~Dorer\orcidlink{0009-0001-1670-5780}}
\addrc
        
\author{M.~Friedl\orcidlink{0000-0002-7420-2559}}
\addrk
        
\author{A.~Garai}
\addrd
        
\author{V.M.~Ghete\orcidlink{0000-0002-9595-6560}}
\addrk
        
\author{D.~Hauff}
\addrd
        
\author{F.~Jeanneau}
\addrb
        
\author{E.~Jericha\orcidlink{0000-0002-8663-0526}}
\addrc
        
\author{M.~Kaznacheeva}
\addra
        
\author{A.~Kinast}
\addra
        
\author{H.~Kluck\orcidlink{0000-0003-3061-3732}}
\addrk
        
\author{A.~Langenk\"{a}mper}
\addrd
        
\author{T.~Lasserre}
\addra
\addrl
        
\author{M.~Mancuso}
\addrd
        
\author{R.~Martin}
\addrb
\addrc
        
\author{A.~Mazzolari}
\addrm
        
\author{E.~Mazzucato}
\addrb
        
\author{H.~Neyrial}
\addrb
        
\author{L.~Oberauer}
\addra
        
\author{T.~Ortmann}
\addra
        
\author{L.~Pattavina}
\addra
\addrn
        
\author{L.~Peters}
\addra
        
\author{F.~Petricca}
\addrd
        
\author{W.~Potzel}
\addra
        
\author{F.~Pr\"{o}bst}
\addrd
        
\author{F.~Pucci}
\addrd
        
\author{F.~Reindl\orcidlink{0000-0003-0151-2174}}
\addrc
\addrk
        
\author{M.~Romagnoni}
\addrm
        
\author{J.~Rothe}
\addra
        
\author{N.~Schermer\orcidlink{0009-0004-4213-5154}}
\addra
        
\author{J.~Schieck}
\addrc
\addrk
        
\author{S.~Sch\"{o}nert}
\addra
        
\author{C.~Schwertner}
\addrc
\addrk
        
\author{L.~Scola}
\addrb
        
\author{G.~Soum-Sidikov\orcidlink{0000-0003-1900-1794}}
\addrb
        
\author{L.~Stodolsky}
\addrd
        
\author{M.~Tamisari}
\addrm
\addro
        
\author{C.~Tomei}
\addrf
        
\author{M.~Vignati\orcidlink{0000-0002-8945-1128}}
\addrf
\addrj

\begin{abstract}
Rare-event search experiments located on-surface, such as short-baseline reactor neutrino experiments, are often limited by muon-induced background events. Highly efficient muon vetos are essential to reduce the detector background and to reach the sensitivity goals. We demonstrate the feasibility of deploying organic plastic scintillators at sub-Kelvin temperatures. For the NUCLEUS experiment, we developed a cryogenic muon veto equipped with wavelength shifting fibers and a silicon photo multiplier operating inside a dilution refrigerator. The achievable compactness of cryostat-internal integration is a key factor in keeping the muon rate to a minimum while maximizing coverage. The thermal and light output properties of a plastic scintillation detector were examined. We report first data on the thermal conductivity and heat capacity of the polystyrene-based scintillator UPS-923A over a wide range of temperatures extending below one Kelvin. The light output was measured down to 0.8\,K and observed to increase by a factor of 1.61\,$\pm$\,0.05 compared to 300\,K. The development of an organic plastic scintillation muon veto operating in sub-Kelvin temperature environments opens new perspectives for rare-event searches with cryogenic detectors at sites lacking substantial overburden.
\end{abstract}

\maketitle

%\tableofcontents

\newpage
%%%%%%%%%%%%%%%%%%%%%%%%%%%%%%%%%%%%%%%%%%%
%%%%%%%%%%%%% Introduction %%%%%%%%%%%%%%%%
	\section{Introduction}

Experiments searching for rare events - such as coherent elastic neutrino-nucleus scattering (CE$\nu$NS) \cite{akimov2017observation,akimov2021first,Singh:2017jow,SalagnacM7:2020,Hakenmuller:2019ecb,Aguilar:2019jlr,Agnolet:2016zir,Akimov:2019ogx,Belov:2015ufh} or direct dark matter interactions \cite{abdelhameed2019first,alkhatib2021light,armengaud2019searching} - are critically dependent on attaining the lowest possible level of background radiation in the target detectors. In particular, cosmic-ray muons can constitute a significant background to the signal events, either directly through energy deposition in the target detectors or indirectly through e.g.\ muon-induced spallation neutrons \cite{heusser1995low}. To mitigate muon-induced background events, rare-event search experiments are usually deployed in laboratory facilities deep underground. In the case of the search for CE$\nu$NS, on the contrary, experiments are commonly located in the immediate vicinity of intense artificial neutrino sources. For instance, the low energy (MeV) of reactor antineutrinos together with their large flux make nuclear fission reactors favorable neutrino sources for a measurement of elastic neutrino-nucleus scattering in the fully coherent regime. Artificial neutrino sources are typically located near the surface, where cosmic-ray muons are of particular concern. Experiments conducted at sites with shallow overburden ($<$\,10\,m.w.e.) hence rely on the deployment of sophisticated shielding as primary method to reduce background radiation. Highly efficient muon vetos are thus essential to render rare-event searches in on-surface laboratories possible.

\vspace{3em}
Many rare-event search experiments, such as RICOCHET \cite{SalagnacM7:2020}, MINER \cite{Agnolet:2016zir} or NUCLEUS \cite{Angloher:2019flc}, are based on cryogenic detectors, since they exhibit excellent energy resolution. This entails the need for low temperatures of $\mathcal{O}$(10\,mK) and hence complex cryogenic facilities. State-of-the-art dilution refrigerators \cite{BF_LD,CC_Cryo,Oxford_Cryo} consist of vertical cylindrical apparatuses in which the multi-stage cryogenic infrastructure occupies the space above the experimental volume. Such construction makes hermetic 4\,$\pi$ shielding of the cryogenic detectors a technical challenge. Since atmospheric muons characteristically follow a cos$^2\theta$ angular zenithal distribution, it is crucial to minimize any coverage gap in the muon veto, especially directly above the target detector. Besides, the typically slow time response of cryogenic detectors ($\mathcal{O}$(100\,$\mu$s)) poses stringent constraints on the tolerable anti-coincident muon rate. In order to minimize the induced target detector dead time, it is essential to keep the muon trigger rate to a minimum. This, in turn, requires a compromise between a highly efficient, yet at the same time compact muon veto and favors a solution that is directly implemented in the interior of the cryostat, also considering the large experimental volume offered by new generation dilution refrigerators. The achievable compactness of cryostat-internal muon veto solutions compared to necessarily vast cryostat-external alternatives is a key advantage. Deploying organic plastic scintillators at temperatures of $\mathcal{O}$(100\,mK) is not trivial since it requires thorough investigation of the detectors’ thermal properties and, accordingly, tailored solutions for thermalization. Moreover, operating organic plastic scintillators inside a running dilution refrigerator necessitates a sophisticated concept for light read-out within the cryogenic infrastructure.

In this paper, we demonstrate the feasibility of instrumenting and operating organic plastic scintillators for particle detection in sub-Kelvin temperature environments. In comparison to conventional cryogenic detectors, plastic scintillator detectors exhibit much faster time response ($\mathcal{O}$(100\,ns)) and can be scaled up very cost-effectively to large kilogram-scale volumes without compromising detector performance. We have developed a novel cryogenic muon veto for the NUCLEUS experiment: a disk-shape plastic scintillator operated inside a cryostat at temperatures of $\sim$\,850\,mK. The muon veto disk is instrumented with wavelength shifting (WLS) fibers and a silicon photo multiplier (SiPM) operated at 300\,K to collect and detect the scintillation light. The paper is organized as follows. Section \ref{sec:lowT} evaluates the feasibility of deploying organic plastic scintillators at sub-Kelvin temperatures: we describe (1) the thermal properties (i.e., thermal conductivity and heat capacity) and (2) the temperature dependence of the measured light signals of a polystyrene-based scintillation detector. In section \ref{sec:NUCLEUSCMV} we present the first sub-Kelvin commissioning of a cryogenic muon veto inside a running dry dilution refrigerator. An evaluation of the detector performance of a cryogenic muon veto for the NUCLEUS experiment is presented.

%%%%%%%%%%%%%%%%%%%%%%%%%%%%%%%%%%%%%%%%%%%
%%%%%%%%%%%%% Section 2 %%%%%%%%%%%%%%%%%%%

    \section{Organic Plastic Scintillators at Sub-Kelvin Temperatures}
	\label{sec:lowT}
 
To operate organic plastic scintillator within the experimental volume of a $^3$He\,/\,$^4$He dilution refrigerator, the detector’s thermal material and light output properties down to sub-Kelvin temperatures are of central interest. 

The essential intrinsic properties of a material governing its thermalization behavior are the thermal conductivity and the heat capacity: the thermal conductivity determines the efficiency of heat transfer through conduction; the heat capacity influences thermal energy storage and release. Generally, achieving efficient thermalization of amorphous polymers below 1\,K within a reasonable timeframe can be challenging due to their inherently low thermal conductivity and considerable heat capacity \cite{ventura2010art}, impeding efficient heat transfer and requiring substantial energy removal. Additionally, the thermal conductivity and heat capacity of amorphous polymers are known to differ by up to one order of magnitude between different polymers at similar temperatures \cite{bradley2013properties}. Therefore, in order to develop tailored cool-down schemes, precise knowledge of the values and temperature dependencies of these parameters is crucial. 

The effect of temperature on the photoluminescence properties of organic plastic scintillators has been studied most commonly in the 0\,$^{\circ}$C to the +50\,$^{\circ}$C temperature range \cite{beddar2012possible,peralta2018temperature,buranurak2013temperature}. In particular, the time response of organic plastic scintillators, which is one crucial parameter for efficient background suppression by a particle veto system, is known to be affected by temperature. Notably, the scintillation decay time of various organic scintillators has been observed to increase towards lower temperatures \cite{sorensen2018temperature}.

For either property, there is scarcity of reported studies conducted in the lower temperature range of $\mathcal{O}$(mK - 1\,K). In the following, we present measurements of the thermal properties and the temperature dependence of the measured light signals of a plastic scintillation detector down to sub-Kelvin temperatures. The organic plastic scintillator under study was the polystyrene-based scintillator UPS-923A \cite{UPS}, which is doped with 2\,\% PTP and 0.03\,\% POPOP.

%%%%%%%%%%%%%%%%%%%%%%%%%%%%%%%%%%%%%%%%%%%
%%%%%%%%%%%%% Section 2.1 %%%%%%%%%%%%%%%%%

	\subsection{Thermal Material Properties of a Polystyrene-based Scintillator}
	\label{sec:therma}
 
\begin{figure}[htbp]
\begin{center}
\includegraphics[width=.8\linewidth]{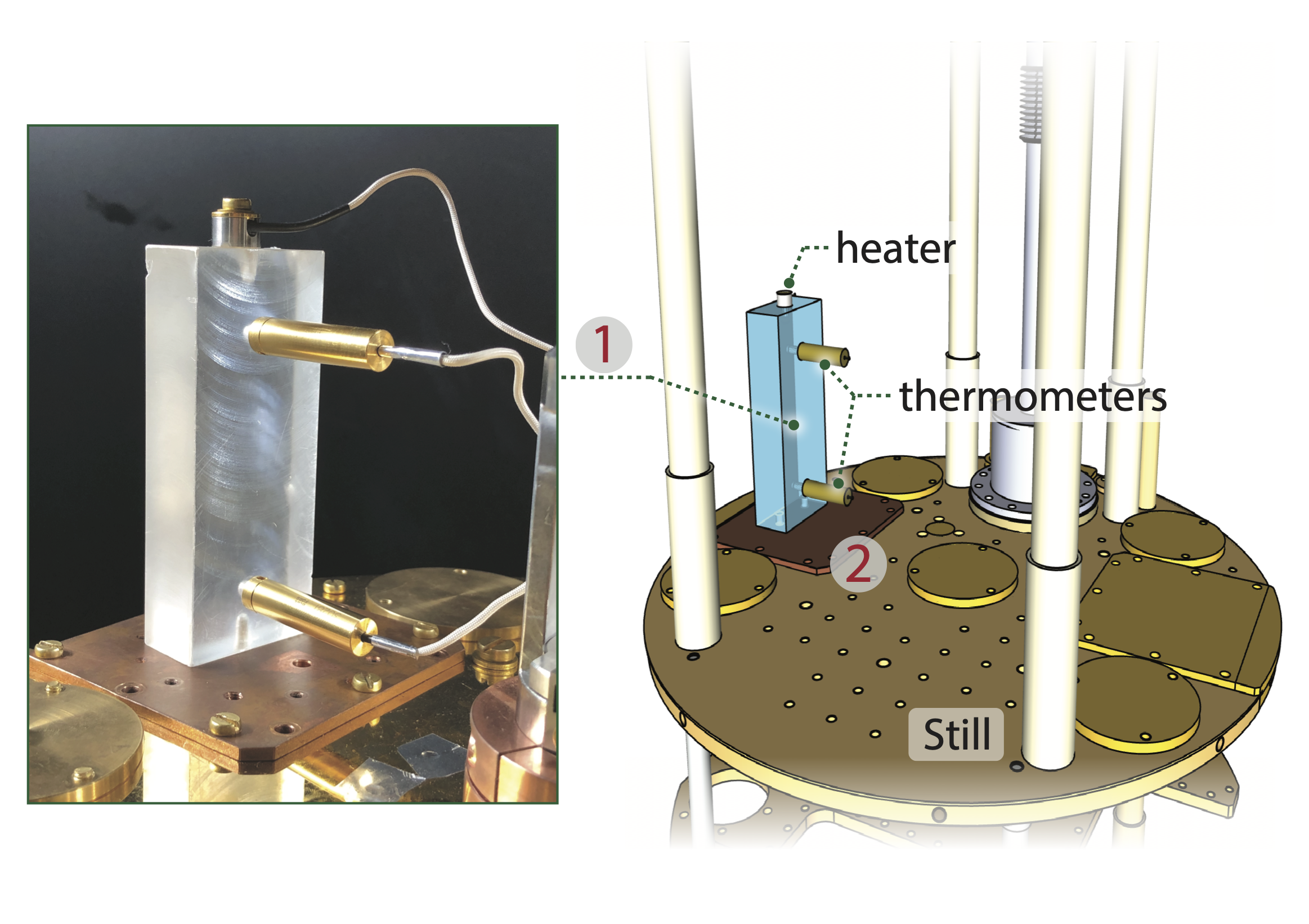}
\caption{Picture of the plastic scintillator sample installed in the dry dilution refrigerator. The rectangular plastic scintillator (1) is thermally coupled (here) to the still stage (at $\sim$\,850\,mK) through a customized copper mounting plate (2) thermalizing it via the contact pressure from the bottom. Two thermometers are screwed in the sample and a heater to its top.}
\label{fig_setup_TC}
\end{center}
\end{figure}

We instrumented a 100\,mm\,$\times$\,40\,mm\,$\times$\,20\,mm rectangular sample of the plastic scintillator UPS-923A, mount\-ed inside a dry dilution refrigerator, to measure its thermal conductivity and heat capacity. The experimental setup, depicted in figure \ref{fig_setup_TC}, consists of the sample with a heating element affixed to the top and two thermometers positioned 65\,mm apart on the side. The sample is connected to a copper mounting plate, which can be attached to any stage of the cryostat with excellent thermal contact (refer to figure \ref{fig_concept} for cryostat stage details), thermalizing the sample via contact pressure. To ensure optimal thermal contact between the cryostat stage and the sample's bottom surface, a thin layer of Apiezon grease \cite{Apiezon} is applied. Additionally, the mounting plate holds supplementary heater and thermometer, enabling both two-probe and four-probe measurements. \\

For the measurement of the thermal conductivity, the "steady longitudinal flow" technique from reference \cite{ventura2010art} was used. By applying a constant power to the heater attached to top of the sample, a stable temperature gradient arises along the length of the sample. The relation between the supplied power $P$ and the thermal conductivity $k$ is given by:

\begin{equation}
    \label{eq:longitudinal_flow}
    P = -\frac{A}{L} \cdot \int_{T_1}^{T_2} k(T) dT
\end{equation}
with $A$ corresponding to the sample's cross-sectional area and $L$ its length. The temperature at the top of the sample (i.e., close to the heater) is denoted as $T_1$ and the temperature at the bottom of the sample as $T_2$. Depending on the temperature range, the sample's bottom temperature $T_2$ was monitored either for a two-probe measurement with the supplementary thermometer of the heat bath (i.e., the cryostat stage) or for a four-probe measurement directly in the lower end of the sample. For temperatures above 1\,K, the thermal contact resistance between the polystyrene sample and the copper mounting plate is assumed to be negligible \cite{ventura2010art}. Additionally, considering that the thermal conductivity of the polystyrene sample is expected to be several orders of magnitude lower than that of the copper mounting plate, the contribution of the latter can be disregarded. Therefore, it suffices to use the supplementary thermometer of the heat bath to monitor $T_2$ for a two-probe measurement. In this case, the distance $L$ corresponds to the length from the bottom of the sample to the top thermometer. For temperatures below 1\,K, the thermal contact resistance becomes non-negligible \cite{ventura2010art}. To mitigate the effects on the measurement, the supplementary heater in the mounting plate is utilized (instead of relying on the heat bath) to maintain a constant temperature $T_2$ at the lower sample thermometer. In the case of this four-probe measurement, the distance $L$ corresponds to the spacing between the two thermometers. By keeping $T_2$ constant and measuring $T_1$ for different input powers $P$, a $P(T)$ curve (as given by equation \ref{eq:longitudinal_flow}) can be obtained. Its derivative yields the thermal conductivity $k(T)$:
\begin{equation}
    \label{eq:conductivity_from_PT_curve}
    k(T) = \frac{L}{A} \frac{dP}{dT}.
\end{equation}

\begin{figure*}[t!]
    \centering
    \begin{subfigure}{.5\textwidth}
    \centering
      \includegraphics[width=.93\linewidth]{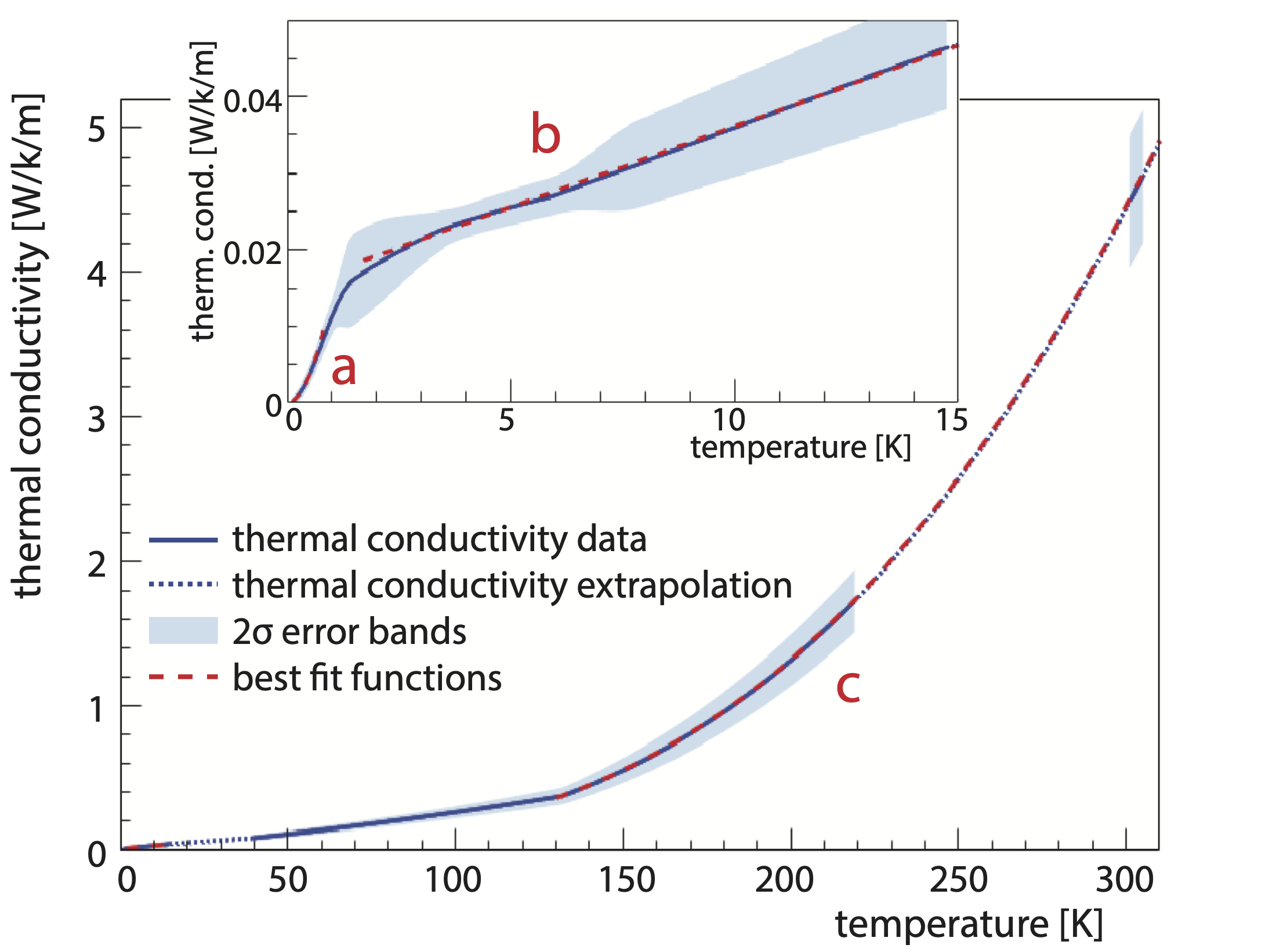}
      \caption{Thermal conductivity plot.}
      \label{fig_plot_TC}
    \end{subfigure}%
    \begin{subfigure}{.5\textwidth}
      \centering
      \includegraphics[width=.93\linewidth]{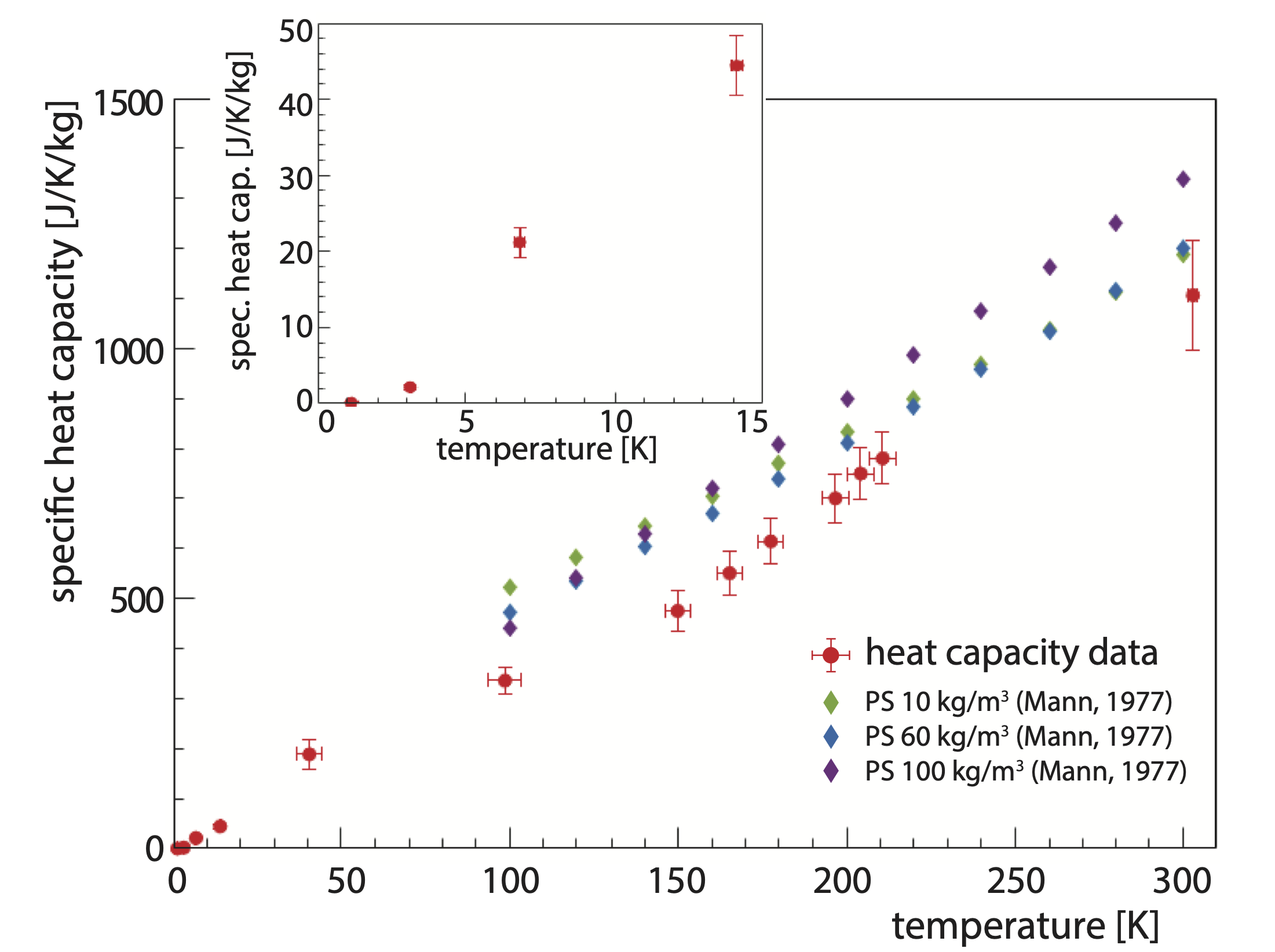}
     \caption{Heat capacity plot.}
      \label{fig_heat_capacity}
    \end{subfigure}
    \caption{($\textbf{a}$) shows the results of the thermal conductivity measurement between sub-Kelvin and room temperature. The error bands correspond to a 2\,$\sigma$ uncertainty. The increased uncertainty in the region between 1K and 3K is due to the unknown thermal contact resistance in the conducted two-probe measurement series. ($\textbf{b}$) shows the obtained values for the heat capacity of UPS-923A. The error bars correspond to a 1\,$\sigma$ uncertainty. Also shown are literature values of polystyrene samples of different densities, taken from reference \cite{mann1977user}. The polystyrene-based UPS-923A has a density of 1060\,kg/m$^3$.}
\end{figure*}

The thermal conductivity of the polystyrene-based scintillator UPS-923A was experimentally determined across the temperature range of 100\,mK to room temperature by measuring the $P(T)$ curve over multiple runs at different cryostat stages. Due to the technical characteristics of the cooling mechanism, it is not possible to maintain arbitrary temperatures in the full range. For this reason, there are no available measuring points between 15\,K - 40\,K and 220\,K - 302\,K. The corresponding results are presented in figure \ref{fig_plot_TC}. The following temperature regions are distinguished:

$i)$ In agreement with Klemens' theory of the thermal conductivity of amorphous solids \cite{klemens1985theory}, we observe a quadratic temperature dependence of the thermal conductivity in the sub-Kelvin region: the best fit between 0.1\,K and 0.7\,K (fit range $a$) is given by a function of the form $k_{a}$\,=\,(0.0146\,$\cdot$\,$T^{1.961}$). 

$ii)$ In the "plateau" region at temperatures of $\mathcal{O}$(1\,K), as suggested in reference \cite{pobell2007matter}, the values for thermal conductivity change only marginally: between 2\,K and 15\,K (fit range $b$), the measured values can be well described by a linear function of the form $k_{b}$\,=\,(0.00211\,$\cdot$\,$T+\,0.01501$). As a benchmark, at 4\,K, the thermal conductivity of UPS-923A was found to be $k$($4\,K$)\,=\,(0.02371\,$\pm$ 0.00107)\,W/K/m, whereas the thermal conductivity of copper is quoted in literature as $k$($4\,K$)\,=\,320.4\,W/K/m at this temperature \cite{bradley2013properties}. 

$iii)$ Above 100\,K up to room temperature, the thermal conductivity shows cubic temperature dependence: the extrapolation up to room temperature, best described by a function of the form $k_{c}$\,=\,(1.65\,$\cdot$\,$10^{-7}$\,$\cdot$\,$T^3$), is based on the measured $k$-values between 130\,K and 220\,K (fit range $c$) and is found to be in good agreement with the measured value at 302\,K. 

To estimate the systematic uncertainty on the thermal conductivity $k$, we take into account a fixed uncertainty of 5\,\% on the sample mass, the geometrical factor $g\,=\,A/L$ and the thermometer calibration. The uncertainty on the set heater power value is estimated to be below 1\,\%. Additional statistical uncertainties have been extracted from the fitting parameters of the $P(T)$ curve. The increased uncertainty of $k(T)$ between 1\,K and 3\,K accounts for the non-negligible Kapitza resistance at these temperatures and has been evaluated from a comparison of two-probe and four-probe data. The measured thermal conductivity values of the polystyrene-based UPS-923A are consistent with the expected behavior of an amorphous polymer at low temperatures, comparable to other polymers such as Nylon or Kapton \cite{bradley2013properties}. However, as the temperature approaches room temperature, the thermal conductivity of UPS-923A is found to be significantly higher.

From the measured values of the plastic scintillator's thermal conductivity $k(T)$, its heat capacity $C(T)$ can be determined according to \cite{ventura2010art}:
\begin{equation}
    \label{eq:heat_capacity}
    C(T) = \frac{g\cdot k(T) \cdot \tau(T)}{m}
\end{equation}
with $m$ denoting the sample mass and $g$ its geometrical factor. $\tau$ represents the time constant of the exponential decline of temperature back to base temperature, once the sample heater is switched off. To get an estimate of the heat capacity of UPS-923A, the time constant $\tau$ was recorded at various temperatures between 1\,K and 300\,K. The resulting heat capacity values are displayed in figure \ref{fig_heat_capacity}. The heat capacity values approximately follow a $T^3$ dependency at temperatures below 10\,K and become increasingly linear towards higher temperatures. As a benchmark, at 3\,K, the specific heat capacity of UPS-923A was found to be $C$($3\,K$)\,=\,(2.173\,$\pm$\,0.348)\,J/K/kg, whereas the specific heat capacity of copper is quoted in literature as $C$($3\,K$)\,$<$\,0.1\,J/K/kg at this temperature \cite{bradley2013properties}. The uncertainty on the heat capacity $C$ is mainly propagated from the uncertainty on the thermal conductivity $k$ – this contributes a 5\,\% to 15\,\% uncertainty depending on the different data points. Smaller uncertainty contributions originate from the uncertainties on the base temperature $T_0$ ($<$\,3\,\%) and from the fit of the time constant $\tau$ ($<$\,1.2\,\%). The obtained results are in agreement with the expected temperature dependence of the specific heat capacity of amorphous solids, as reported in reference \cite{ventura2010art}. Additionally, the observed linear behaviour above 100\,K is consistent with the previously reported findings on pure polystyrene with varying densities, as documented in reference \cite{mann1977user}.

\begin{figure}[htbp]
\begin{center}
\includegraphics[width=.93\linewidth]{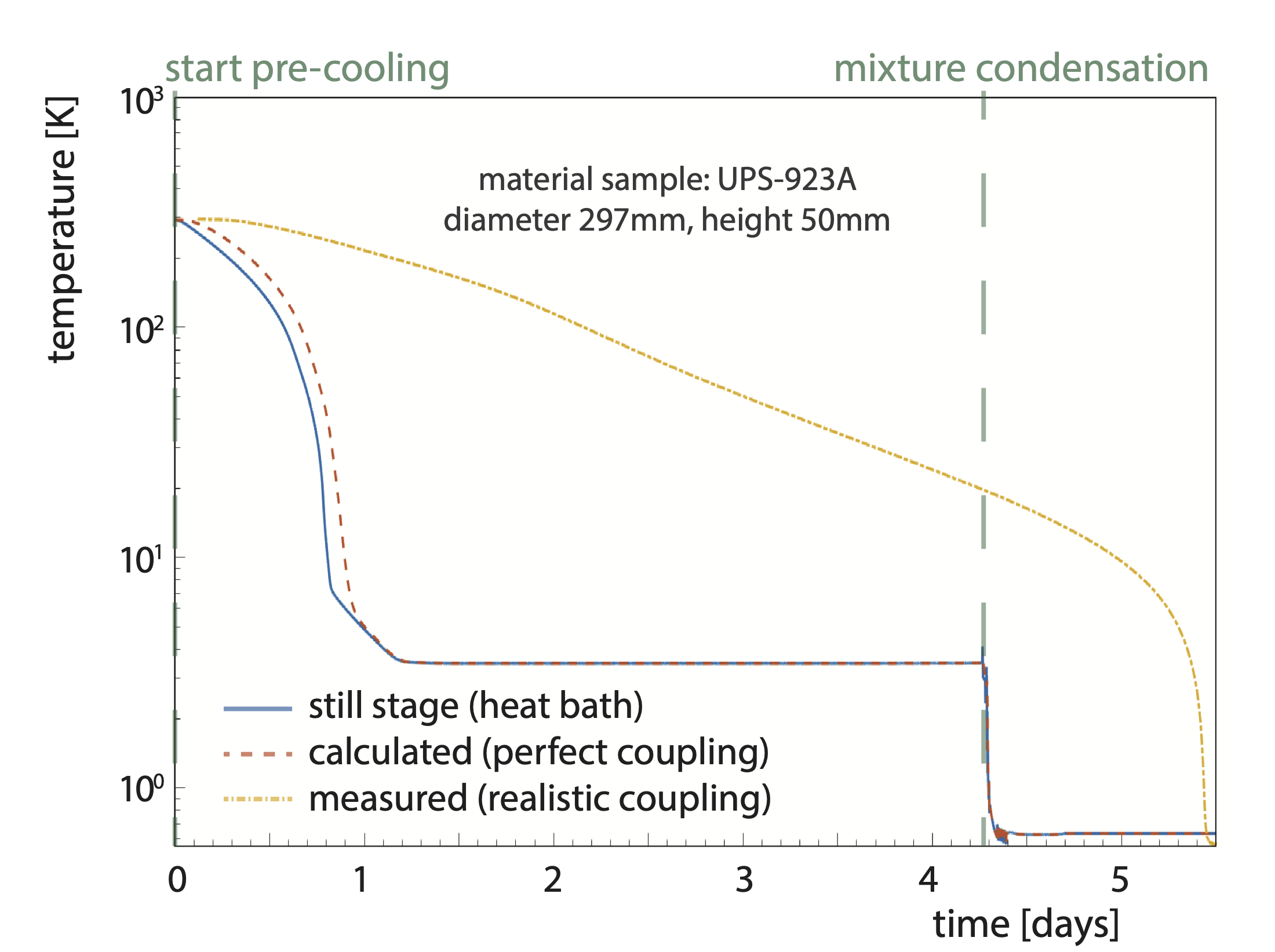}
\caption{Cool-down curves of the plastic scintillator in logarithmic temperature scale. The orange curve represents the simulated cool-down curve assuming perfect thermal coupling. The yellow curve represents the measured cool-down curve with a tailored thermalization concept described in detail in section \ref{sec:NUCLEUSCMV}. The blue curve shows the reference cool-down curve of the still stage. The start of pre-cooling and the condensation of the $^3$He\,/\,$^4$He mixture to reach base temperature are indicated (\textit{green}).}
\label{fig_cool-down}
\end{center}
\end{figure}

The obtained thermal properties allow to calculate the ideal cooling time of a sample of the plastic scintillator UPS-923A. In this study, one side of a cylindrical plastic scintillator sample, measuring 297 mm in diameter and 50 mm in height (similar in dimension to the NUCLEUS cryogenic muon veto developed and commissioned in section \ref{sec:NUCLEUSCMV}), is assumed to be perfectly coupled to the still stage of a cryostat. The entire cooling power of the still stage of $\mathcal{O}$(1\,mW) \cite{BF_LD} is hypothetically considered to be available for thermalizing the plastic.  Using actual cool-down data for the still temperature as boundary condition, we estimated the temperature curve of the opposite side of the plastic scintillator disk. The still stage cool-down curve and the calculated cool-down curve of the plastic scintillator are presented in figure \ref{fig_cool-down}. In this hypothetical case of ideal thermal coupling to a given heat bath, cool-down time is only limited by the intrinsic thermal conductivity and heat capacity: the non-coupled end of the plastic scintillator closely follows the still temperature with a delay of approximately 2\,hours. However, achieving the idealized assumption of perfect coupling over the entire side of the disk is difficult to realize in an actual thermalization experiment. When thermalizing a kilogram-scale piece of an amorphous plastic, practical challenges regarding the realization of the thermal coupling turn out the primary limitation of the achievable cool-down time. These challenges include imperfect surface flatness, limited contact pressure that can be exerted on an amorphous polymer compared to a metal, increasing Kapitza resistance at lower temperatures and considerable differences in thermal contraction coefficients between polystyrene and copper. These factors, all of which are ignored in the ideal case, are severely restricting thermal coupling strength. A sophisticated thermalization concept, taking into account the limiting factors in the implementation of a realistic thermal coupling, was developed and commissioned in section \ref{sec:NUCLEUSCMV} for a cryogenic muon veto for the NUCLEUS experiment. The resulting cool-down curve measured for a prototype plastic scintillator sample (297\,mm in diameter and 50\,mm in height) is likewise depicted in figure \ref{fig_cool-down} and is, as expected, considerably slower than the ideal case.

%This simple case study demonstrates the principal feasibility of cooling several kilograms of a polystyrene-based plastic scintillator to sub-Kelvin operating temperatures within reasonable timeframes, enabling integration into the cryogenic environment of $^3$He\,/\,$^4$He dilution refrigerators. The main difficulty is the realization of the thermal coupling to the cooling capacity of the cryostat. However, as we demonstrate in section \ref{sec:NUCLEUSCMV}, cool-down times of a few days can be achieved with a tailored thermalization concept.

%%%%%%%%%%%%%%%%%%%%%%%%%%%%%%%%%%%%%%%%%%%
%%%%%%%%%%%%% Section 2.2 %%%%%%%%%%%%%%%%%

    \subsection{Temperature Dependence of the Measured Light Signals}

\begin{figure}[htbp]
\begin{center}
\includegraphics[width=.8\linewidth]{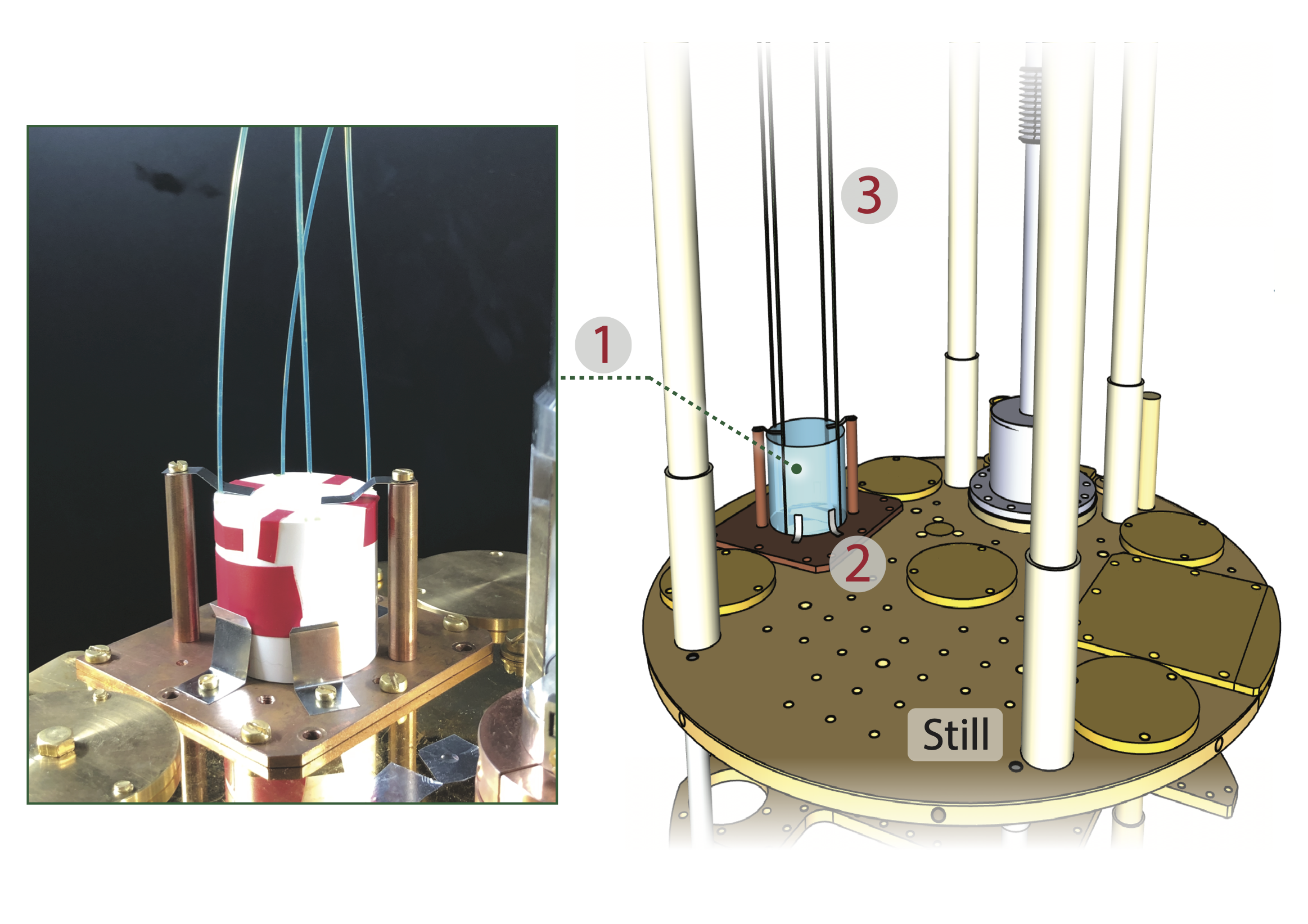}
\caption{Picture of the plastic scintillator sample installed in the dry dilution refrigerator. The cylindrical plastic scintillator (1) is thermally coupled to the still stage (at $\sim$\,850\,mK) through a customized copper mounting plate (2) thermalizing it via the contact pressure from the bottom. The wavelength shifting fibers (3) are guided through the cryostat to the SiPM, which is attached to the bottom of the 300 K plate.}
\label{fig_setup_LY}
\end{center}
\end{figure}

We instrumented a 42\,mm\,($\diameter$)\,$\times$\,42\,mm\,(h) cylindrical sample of the plastic scintillator UPS-923A, mounted on top of the still stage of a dry dilution refrigerator, for light-read out. Four BCF-91A wavelength shifting fibers \cite{SaintGobain2} have been glued into small grooves on the side to collect and guide the scintillation light. The plastic scintillator was wrapped in diffusive polyester foil (Lumirror E6SR from Toray \cite{TORAY}) to increase light collection. A KETEK PE3325-WB TIA TP SiPM module \cite{KETEK} attached to the bottom of the 300K plate of the cryostat was used to detect the scintillation light. SiPMs are solid-state single-photon-sensitive devices composed of arrays of independent avalanche photo diodes. When operated in Geiger mode, single-photons can trigger a self-sustained process of avalanche ionization, which is converted into a macroscopic current flow. The signals were acquired with a self-triggering Struck SIS3316 analog-to-digital converter (FADC) \cite{Struck}. The experimental setup is depicted in figure \ref{fig_setup_LY}. Data with the scintillation detector have been taken in different phases of a cool-down cycle: i) in thermalized condition at 4\,K at the end of the pulse tube pre-cooling; ii) at the still stage base temperature of 800\,mK; iii) after warm-up to room temperature at 300\,K, as reference.

\begin{figure}[htbp]
\begin{center}
\includegraphics[width=.93\linewidth]{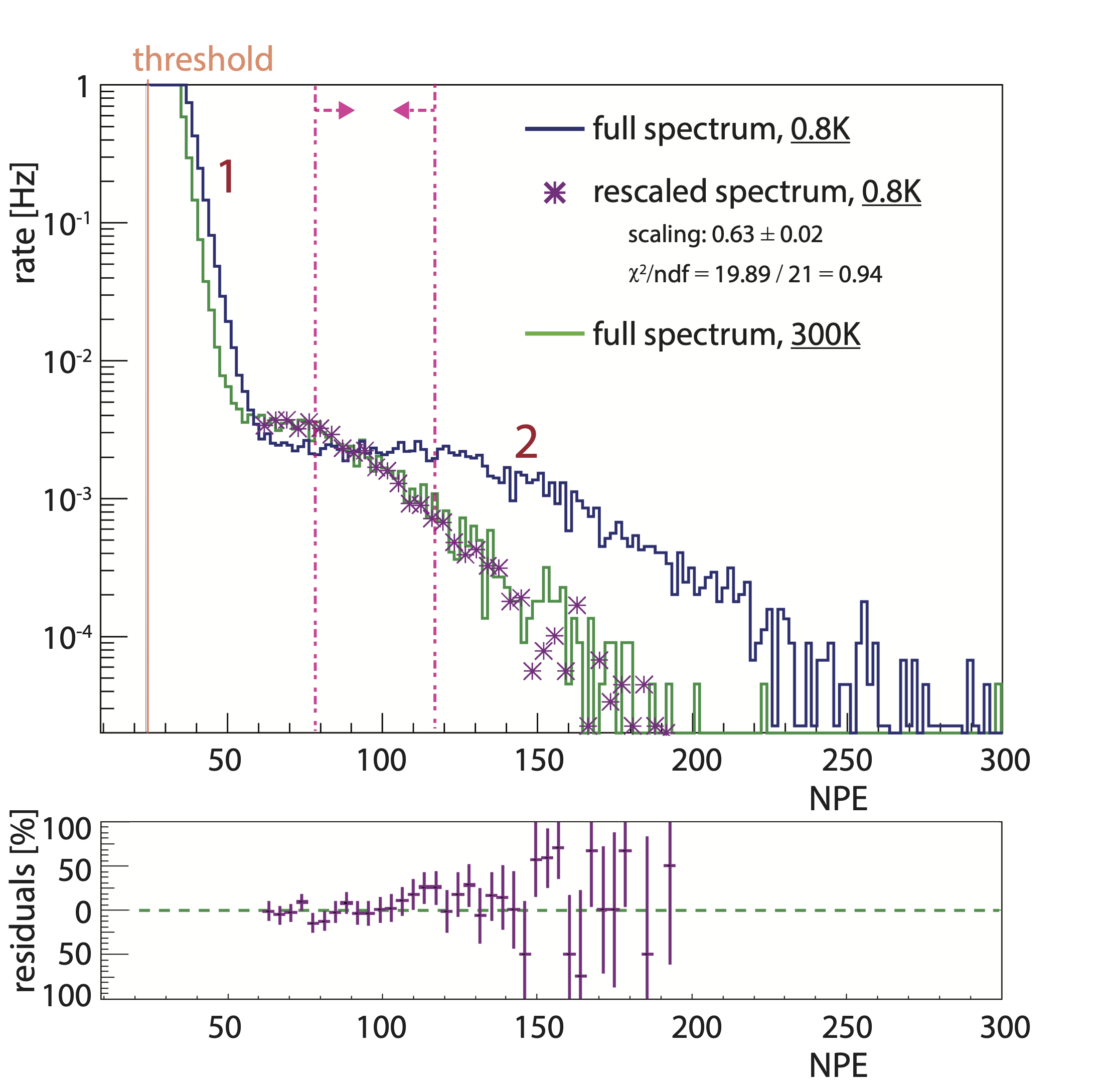}
\caption{Gain-calibrated NPE spectra of events recorded with the scintillation detector thermalized at 800\,mK (\textit{blue}) and at 300\,K (\textit{green}). Both spectra feature (1) a region for low energetic background events cut at a certain threshold (\textit{orange}) and (2) a continuous spectrum at higher energies corresponding to muon events. The range between 80\,PE and 120\,PE, taken to build an averaged muon pulse at 0.8\,K (300\,K), is marked (\textit{magenta arrows}) in the NPE spectrum. The best-fit rescaled 800\,mK spectrum (\textit{purple}) is also shown, together with the residuals of the spectra. Error bars are statistical only.}
\label{fig_spectra_LY}
\end{center}
\end{figure}

\begin{figure}[htbp]
\begin{center}
\includegraphics[width=.93\linewidth]{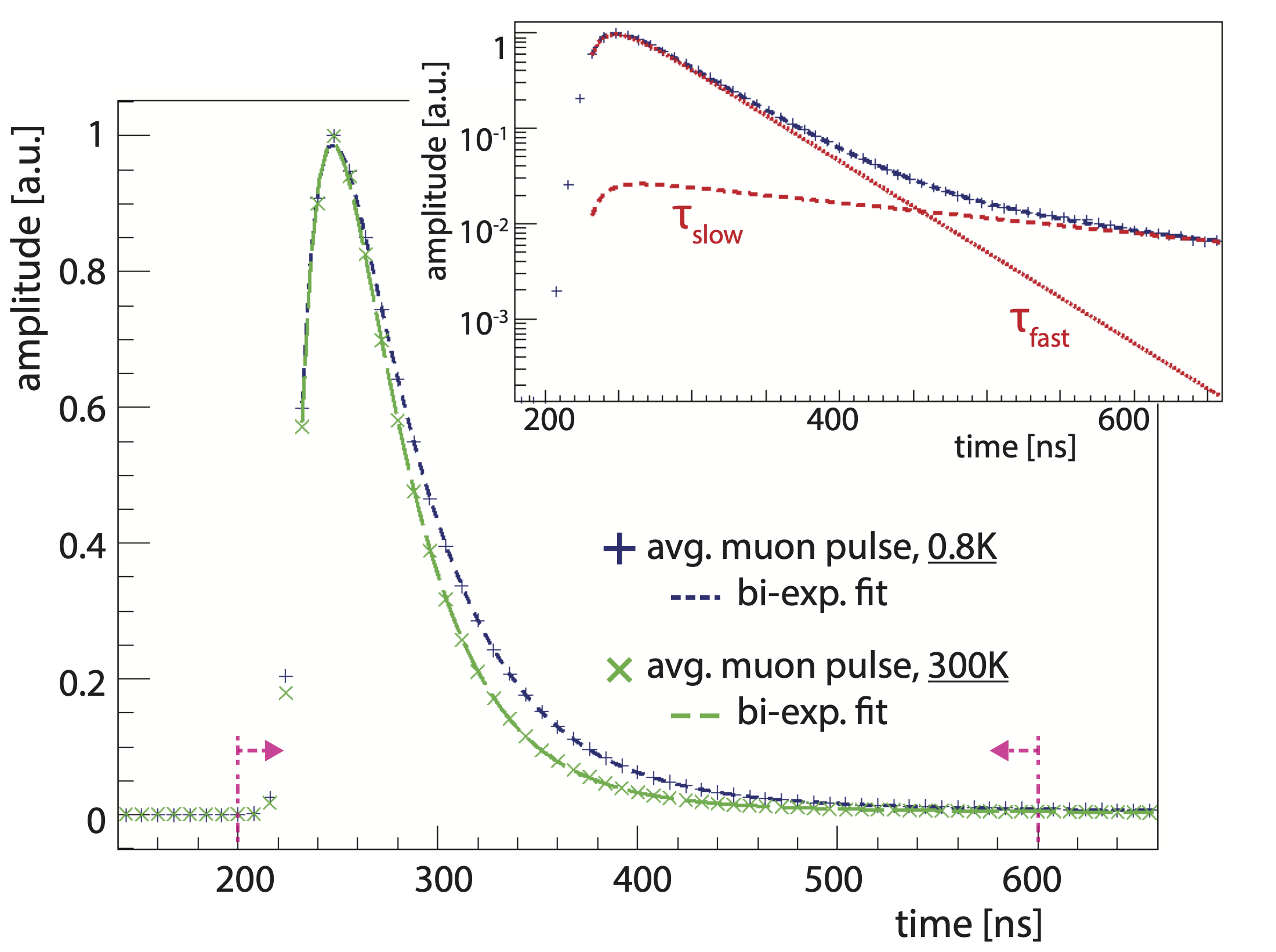}
\caption{Averaged muon pulses captured with the scintillation detector thermalized at 800\,mK (\textit{blue}) and at 300\,K (\textit{green}), together with the respectively applied multi-exponential fits (\textit{dashed lines}). The inset shows the averaged muon pulse at 800\,mK plotted in logarithmic amplitude scale and details individually the fits of the slow and fast decay components of the overall pulse. For better comparability, the pulse amplitude is normalized to 1. The integration gate to determine the pulse charge $Q$ is marked (\textit{magenta arrows}).}
\label{fig_fit}
\end{center}
\end{figure}

\vspace{3em}
The main observable is the integrated charge $\textit{Q}$ of the signal pulse, which is a measure for the collected scintillation light and hence an estimate of the deposited energy by a traversing particle in the plastic scintillator. Figure \ref{fig_spectra_LY} shows the charge spectra of events measured at 0.8\,K and at 300\,K. The value of the pulse charge $\textit{Q}$ can be retrieved from integration of the signal pulses over an interval of 400\,ns (the integration gate is indicated in $\textit{magenta}$ in the muon pulse trace in figure \ref{fig_fit}). The spectrum features (1) an external gamma-background increasing towards threshold and (2) a continuous spectrum reaching to higher charges corresponding to muon events. As illustrated in reference \cite{Wagner_2022}, the large dark count rate of the SiPMs can be used to monitor their stability and calibrate their gain (i.e. the amount of charge created for each detected photon). Charge spectra of dark counts have been recorded for each respective measurement, from which the gain of individual Geiger discharges can be obtained. In analogy to PMTs, the term "number of photoelectrons" (NPE) is used in the following. Since the SiPM is thermally coupled to the 300\,K stage of the cryostat, it is not expected to be subject to significant temperature fluctuations. However, even small ambient temperature changes of a few Kelvin (e.g. due to heat radiation from colder cryostat stages or the working pulse tube, conduction through the adjacent fibers or fluctuations in the ambient temperature in the laboratory) can influence the gain of the SiPM. The calibration of the measured spectra accounts for such possible gain variations of the SiPM and hence allows for a direct comparison of measurements with the plastic scintillator thermalized at different temperatures. In the case of the above measurements at 800\,mK and 300\,K, the gain was determined to be (8346\,$\pm$\,37)\,FADC\,a.u.\,/\,PE (300\,K) and (7284\,$\pm$\,70)\,FADC\,a.u.\,/\,PE (800\,mK), respectively. To quantify the light output of the scintillation detector, we fit a horizontal scaling factor on the spectrum recorded at 800\,mK (\textit{blue}) to bring it in agreement with the reference spectrum recorded at room temperature (\textit{green}). This approach, while not depending on a specific shape model, yields a global light output parameter to interpret the difference between the two spectra. For the best-fit scaling factor of 0.63\,$\pm$\,0.02, a very good agreement is observed between the rescaled 800\,mK spectrum and the 300\,K spectrum, quantified by a $\chi ^2$/ndf of 19.89\,/\,21\,=\,0.94. The resulting rescaled 800\,mK spectrum is shown in figure \ref{fig_spectra_LY} (in \textit{purple}) together with the residuals. In order to not dominate the statistical weight of the fit by the low energy background, the fit range been has been restricted to the muon tail between 70\,PE and 150\,PE. The retrieved best-fit scaling factor corresponds to an increase of the detector’s light output by a factor of 1.61\,$\pm$\,0.05 at 0.8\,K compared to the reference measurement at 300\,K. At 4\,K, the light output is consistent with the measurement at 0.8\,K. Since all measurements at the various temperatures have the same calibrated spectrum of atmospheric muons as their basis, the observed increase in measured NPE implies that, for a given energy deposition in the plastic scintillator, more light is ultimately reaching the SiPM. In a separate run, during the 12\,h warm-up from sub-Kelvin temperature to room temperature, it was also observed that the change in light output continuously follows the change in temperature over a wide range. 

The observed increase of the integrated pulse charge $\textit{Q}$ towards lower temperatures is consistent with what is generally reported at higher temperatures \cite{beddar2012possible,peralta2018temperature,buranurak2013temperature,sorensen2018temperature,homma1987effect,xia2014temperature}: as temperature decreases, photoluminescence quantum yields of organic scintillators are expected to increase, since quenching by non-radiative processes associated with the thermal agitation of organic molecules is expected to decrease \cite{valeur2012molecular}. However, it is important to note here that the light output of the scintillation detector is only an indirect measure of the photoluminescence quantum yield of the organic scintillator due to the intermediate light collection and light guiding processes (e.g. through wavelength shifting fibers and the diffusive polyester foil). Hence, simultaneous temperature-dependent effects occurring after the generation of scintillation light may additionally affect the overall light output of the detector. For instance, a shift of the mean wavelength in the emission spectrum of the plastic scintillator, which has been chosen to match the absorption spectrum of the WLS fibers, might affect the light collection efficiency and thus the measured light output \cite{homma1987effect}.

To investigate a potential temperature-dependency of the time response of the scintillation detector, a standard muon pulse has been obtained at 0.8\,K (4\,K, 300\,K) by averaging over a total of 2239 (2540, 829) single pulses with an integrated charge $\textit{Q}$ between 80\,PE and 120\,PE (the respective charge range is marked ($\textit{magenta}$) in the charge spectrum in figure \ref{fig_spectra_LY}). The resulting standard signal pulse shape can be fitted by a sum of exponential functions \cite{sorensen2018temperature}:

\begin{equation}
\begin{split}
    f(t)=\frac{A_1}{\tau _{fast} - \tau _0}\cdot [exp(-\frac{t-t_0}{\tau _{fast}}) - exp(-\frac{t-t_0}{\tau _{0}})] \\+ \frac{A_2}{\tau _{slow} - \tau _0}\cdot [exp(-\frac{t-t_0}{\tau _{slow}}) - exp(-\frac{t-t_0}{\tau _{0}})]
\end{split}
\end{equation}
with $A_i$ corresponding to the intensities, $\tau_i$ the decay times for the fast and slow light emission components and $\tau_0$ the integration constant of the electronics. The fitted standard muon pulses captured at 0.8\,K and at 300\,K are depicted in figure \ref{fig_fit}.

The values of the fast and slow decay time obtained from the fits of the averaged muon pulses are summarized in table 1 for the measurements with the plastic scintillator thermalized at 0.8\,K, 4\,K and 300\,K.

	\begin{table}[htbp]
	   \centering
	     \caption{Summary of the fast decay time $\tau_{fast}$ and the slow decay time $\tau_{slow}$ of the averaged signal pulses, as well as the decay time $\tau_{SiPM}$ of single-photon SiPM signals acquired at 0.8\,K, 4\,K and 300\,K. The decay times for the fast and slow light emission components have been retrieved from the multi-exponential fit of the averaged muon pulse. Only statistical uncertainties are given.}
	   \begin{tabular}{l|c|c|c}
	        \hline
            & $\tau_{fast}$ [ns] & $\tau_{slow}$ [ns] & $\tau_{SiPM}$ [ns] \\
            \hline
            0.8\,K & 45.4\,$\pm$\,1.4 & 270.6\,$\pm$\,68.4 & 19.0\,$\pm$\,2.1 \\
            4\,K & 46.2\,$\pm$\,1.2 & 343.4\,$\pm$\,111.4 & 18.9\,$\pm$\,1.8 \\
            300\,K & 35.4\,$\pm$\,2.1 & 200.6\,$\pm$\,61.1 & 19.0\,$\pm$\,4.0 \\
	        \hline
	   \end{tabular}
	    \label{tab:performance}
	\end{table}

The fast decay time, $\tau_{fast}$, remains constant within the uncertainty for the measurements at 0.8\,K and 4\,K and decreases by a factor of 1.29\,$\pm$\,0.09 for the measurement at 300\,K. The slow decay time, $\tau_{slow}$, was found not to change within a large uncertainty of $\sim$\,30\,\%, which is entirely statistical in origin. 

In order to monitor the SiPM performance, we conducted an independent cross-check: the averaged pulse shape of single-photon SiPM signals (induced by dark counts) yields information about the time response of the SiPM and has been compared for the measurements at the respective temperatures. The recovery time of the SiPM, $\tau_{SiPM}$, is dominated by the pre-amplifying electronics, which features a bandwidth limit of 12.5\,MHz \cite{KETEK}. It was found to be constant at (18.9\,$\pm$\,1.6)\,ns for all of the above measurements. Hence, the observed temperature dependency of the time response of the measured light signals cannot be attributed to the performance of the SiPM during different phases of the cool-down cycle, but is presumably originating from the processes of scintillation light generation, collection and guidance.

The measured properties of the detector’s light output and time response at temperatures below 1\,K validate the performance of a polystyrene-based scintillator inside a running dry dilution refrigerator. In particular, the pulse decay time in the sub-Kelvin temperature range is only modestly slowed down, making it well-suited for deployment as fast anti-coincident muon veto. Based on that, the following section \ref{sec:NUCLEUSCMV} describes the development and the first sub-Kelvin commissioning of a cryogenic muon veto for the NUCLEUS experiment.

\vspace{3em}

%%%%%%%%%%%%%%%%%%%%%%%%%%%%%%%%%%%%%%%%%%%
%%%%%%%%%%%%% Section 3 %%%%%%%%%%%%%%%%%%%
 
    \section{A Cryogenic Muon Veto for the NUCLEUS Experiment} 
    \label{sec:NUCLEUSCMV}

\begin{figure*}[t!]
\begin{center}
\includegraphics[width=1\linewidth]{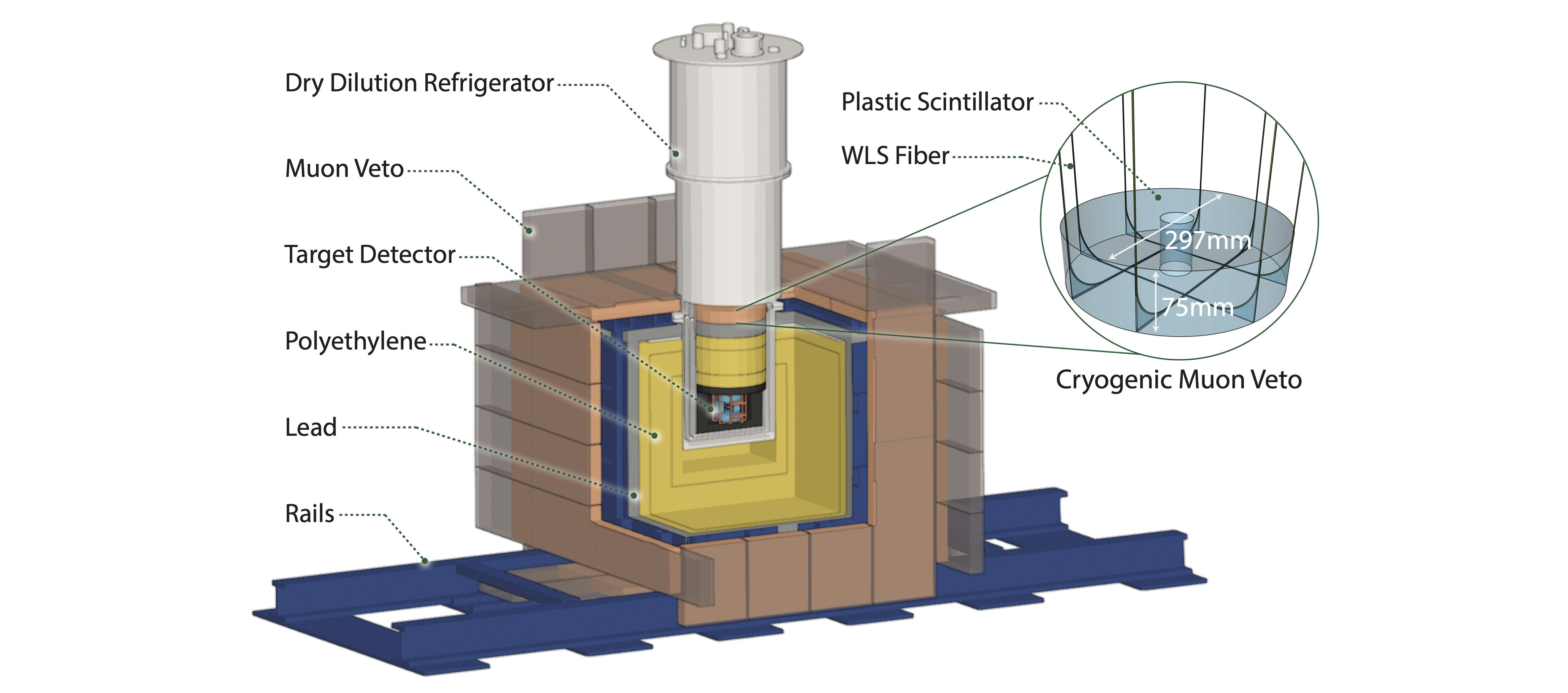}
\caption{Schematic cutaway drawing of the NUCLEUS experimental setup. The NUCLEUS cryostat containing the target detector is shielded by a $\sim$\,1\,m$^3$-sized passive shielding consisting of polyethylene (\textit{yellow}) and lead (\textit{gray}). The shielding is placed on rails in order to access the target detectors. The outermost layer is an active muon veto (\textit{orange}), consisting of 28 single organic plastic scintillator panels each enclosed in a light-tight aluminium box and placed hermetically around the NUCLEUS cryostat. The remaining hole is closed by a cryogenic muon veto (\textit{zoom-in}): an organic plastic scintillator disk mounted in the inside of the cryostat, operating at sub-Kelvin temperatures and instrumented with wavelength shifting fibers.}
\label{fig1}
\end{center}
\end{figure*}

The NUCLEUS experiment will use cryogenic calorimeters inside a dry dilution refrigerator to measure reactor antineutrinos scattering coherently off the target nuclei. The target detector is surrounded by a $\sim$\,1\,m$^3$-sized passive shielding positioned outside the cryostat consisting of polyethylene and low-background lead. Operating at an overburden of 3\,m.w.e., a high muon tagging efficiency especially in the area above the target detectors is crucial to achieve the necessary background reduction to a level of 100\,counts\,/\,(keV$\cdot$kg$\cdot$d). A state-of-the-art muon veto consisting of a compact cube assembly of 28 single organic plastic scintillator panels hermetically placed around the passive shielding has been developed in reference \cite{Wagner_2022}. The arrangement is shown schematically in figure \ref{fig1}.

The operation of an organic plastic scintillation muon veto inside the cryostat of the NUCLEUS experiment at sub-Kelvin temperatures allows to close the remaining hole where the dry dilution refrigerator enters the shielding and hence to increase the overall muon veto efficiency with a minimum of additional active detector material. Extensive GEANT4 \cite{GEANT4:2002zbu} Monte Carlo simulations have been implemented in reference \cite{Goupy_NU22} in order to estimate the residual background level in the NUCLEUS target detector at sub-keV energies. The simulations consider the full NUCLEUS apparatus geometry (including the cryogenic muon veto) and a muon veto anti-coincidence cut at a threshold of 5\,MeV. Preliminary results confirm that an efficiency for muon identification of more than 99\,\% and an associated muon rate of 325\,Hz can be achieved, meeting the requirements of the NUCLEUS experiment. The cryogenic muon veto critically reduces the background of muon-induced events in the target detector's region of interest between 10\,-\,100\,eV by an additional factor of approximately 2\,-\,3 and thus proved to be an essential component for NUCLEUS towards achieving the required benchmark background index of 100\,counts\,/\,(keV$\cdot$kg$\cdot$d).

%%%%%%%%%%%%%%%%%%%%%%%%%%%%%%%%%%%%%%%%%%%
%%%%%%%%%%%%% Section 3.1 %%%%%%%%%%%%%%%%%

	\subsection{Cryogenic Muon Veto Integration into the NUCLEUS Cryostat}
	\label{sec:int}

\begin{figure}[htbp]
\begin{center}
\includegraphics[width=.95\linewidth]{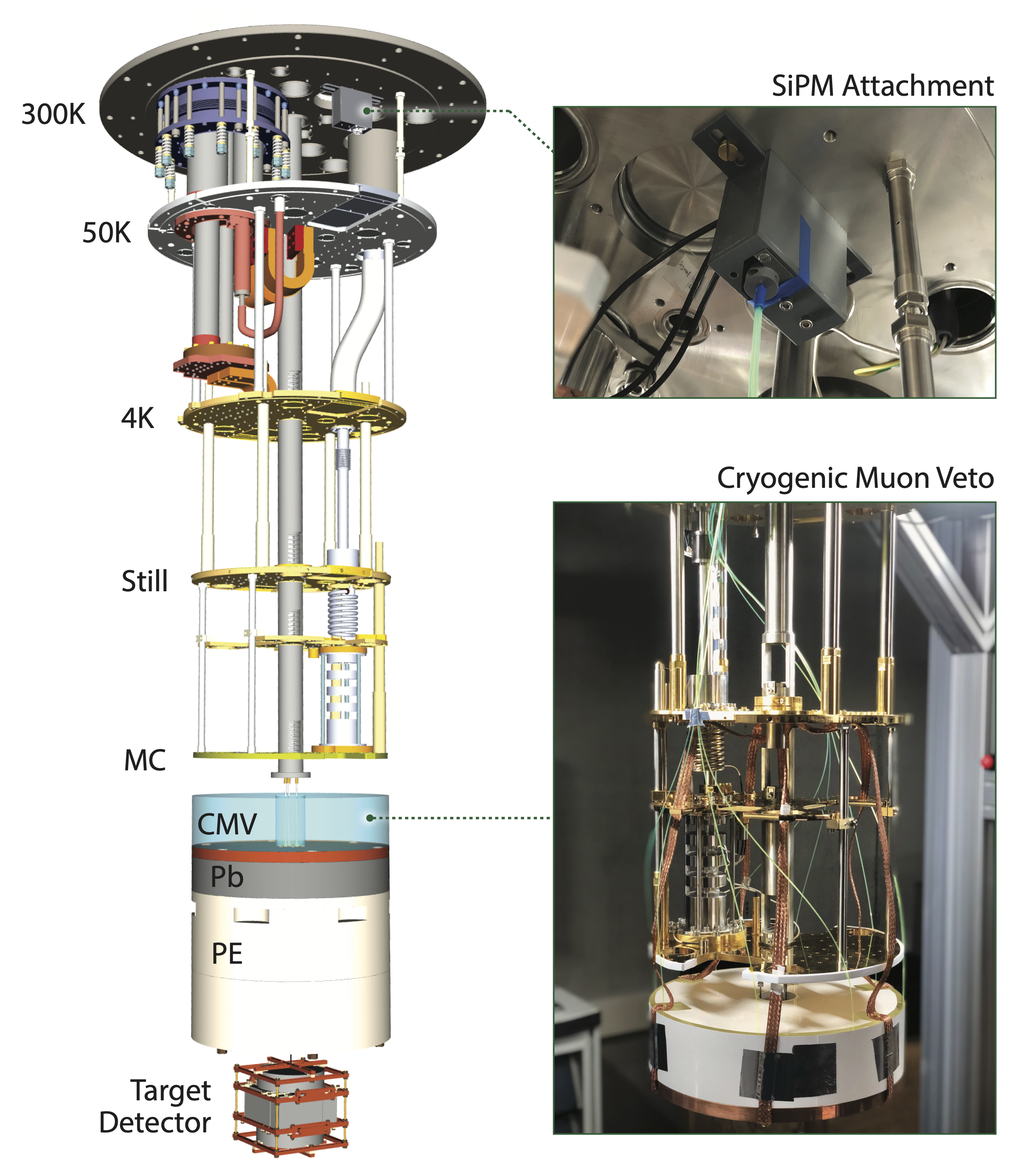}
\caption{Schematic drawing of the NUCLEUS cryogenic muon veto (\textit{CMV}) detector concept with images of the various detector sub-systems, namely the plastic scintillator disk installed underneath the mixing chamber (\textit{MC}) plate of the NUCLEUS dry dilution refrigerator (\textit{bottom right}) and the SiPM attachment to the 300 K plate (\textit{top right}). In addition, the cryostat-internal passive shielding, consisting of a low-background lead disk (\textit{Pb}) and a cylindrical polyethylene block (\textit{PE}), as well as the target detector are indicated.}
\label{fig_concept}
\end{center}
\end{figure}

We have built a cryogenic muon veto specifically for deployment in the NUCLEUS experiment. It consists of a plastic scintillator disk instrumented with a fiber based light-guide system together with a SiPM based read-out system and was commissioned inside a commercial LD\,400 $^3$He\,/\,$^4$He dry dilution refrigerator provided by Bluefors \cite{BF_LD}. The detector concept and the detector integration with the cryogenic infrastructure is shown schematically in figure \ref{fig_concept}. A disk of the plastic scintillator UPS-923A with an outer diameter of 297\,mm and a height of 75\,mm is employed underneath the mixing chamber plate of the NUCLEUS cryostat. At minimum ionization energy loss, a muon deposits about 2 MeV/cm along its passage through a polymer based plastic. Gamma-rays from natural radioactivity, which extend up to 2.6\,MeV in energy, can thus be efficiently discriminated from muons by a simple energy cut. In order to maximize the coverage above the target detector, the outer diameter of the disk was chosen to tightly fit the innermost vessel of the cryostat, leaving a required gap of 4 mm for the feed-through of the detector wiring.

The disk (shown schematically in figure \ref{fig1}) features four bended grooves arranged in a grid pattern, each of which guides a commercially available multi-clad BCF-91A wavelength shifting fiber \cite{SaintGobain2} through the interior of the plastic scintillator, respecting the fibers' minimum recommended bending radius of 5\,cm. Perpendicular grooves are milled at two different depths in order to avoid conflicts due to the crossing of the fibers. The absorption and emission spectra of the fibers match well the transmission spectrum of the UPS-923A plastic scintillator and the range of high photo detection efficiency of the chosen SiPMs. The fibers are glued via the two component clear epoxy resin BC\,600 \cite{cement} into the grooves, assuring thorough fixation and maximization of the contact surface. An inner hole with a diameter of 45\,mm is required for the feed-through of the support structure of the target detector.

The scintillation light produced by the disk is absorbed by the fibers, isotropically re-emitted at a longer wavelength and transmitted through the fibers by total internal reflection. The two ends of each fiber are exiting the grooves towards the top and guided 100\,cm up to the 300\,K plate of the cryostat to be read out on both sides, hence maximizing scintillation light collection. Halfway, the fibers are thermally coupled to the intermediate 4K plate using a copper wire. The large pliability offered by the WLS fibers allows to flexibly guide the light through the cryostat vessels across its stages towards a pre-amplified KETEK PE3325-WB TIA TP SiPM module \cite{KETEK}, operating at an overvoltage of V$_{OV}$\,=\,5\,V. The total of eight ends of the four fibers are bundled to a cross-sectional area of 8\,mm$^2$ by a custom-made PVC connector and attached to the SiPM module, which has a 3\,x\,3\,mm$^2$ active area and a pixel size of 25\,$\mu$m. The connector design ensures precise alignment of the fiber ends parallel to the active area of the SiPM, leaving a safety gap of a few hundred microns between the two surfaces. Thorough optical coupling is ensured with optical grease. To enhance the light collection, the plastic scintillator disk is hermetically covered with the diffusive polyester foil Lumirror E6SR from Toray \cite{TORAY}. The SiPM module is mounted inside a custom-made PVC casing to the bottom of the 300\,K plate inside the cryostat and thus operating at room temperature. Since the time response of SiPMs is significantly delayed with decreasing temperatures due to the built-in quenching resistor \cite{csathy2011development} it was opted against coupling the SiPM to a lower temperature stage of the cryostat. The SiPM signals are fed out of the cryostat via vacuum-tight LEMO feedthroughs into a 16-channel self-triggering Struck SIS3316 FADC with a sampling rate of 125\,MHz \cite{Struck}.

The positioning of a cryogenic muon veto inside the cryostat of the NUCLEUS experiment, directly above the target detectors, requires sub-Kelvin operating temperatures in order to keep the radiative heat load on the lower cryostat sections within tolerable limits. Thermalization of the plastic scintillator at approximately 850\,mK ($T_{still}$) takes advantage of the increased cooling power of the still stage relative to the mixing chamber stage, without, however, significantly increasing the thermal radiation on the detector volume. Moreover, the suitability of the still stage as the thermalization stage for the cryogenic muon veto and the associated operating temperature of $\sim$\,850\,mK has been confirmed by the proof-of-principle measurements in section \ref{sec:lowT} - encompassing thermal material and light output properties.

\begin{figure}[htbp]
\begin{center}
\includegraphics[width=.95\linewidth]{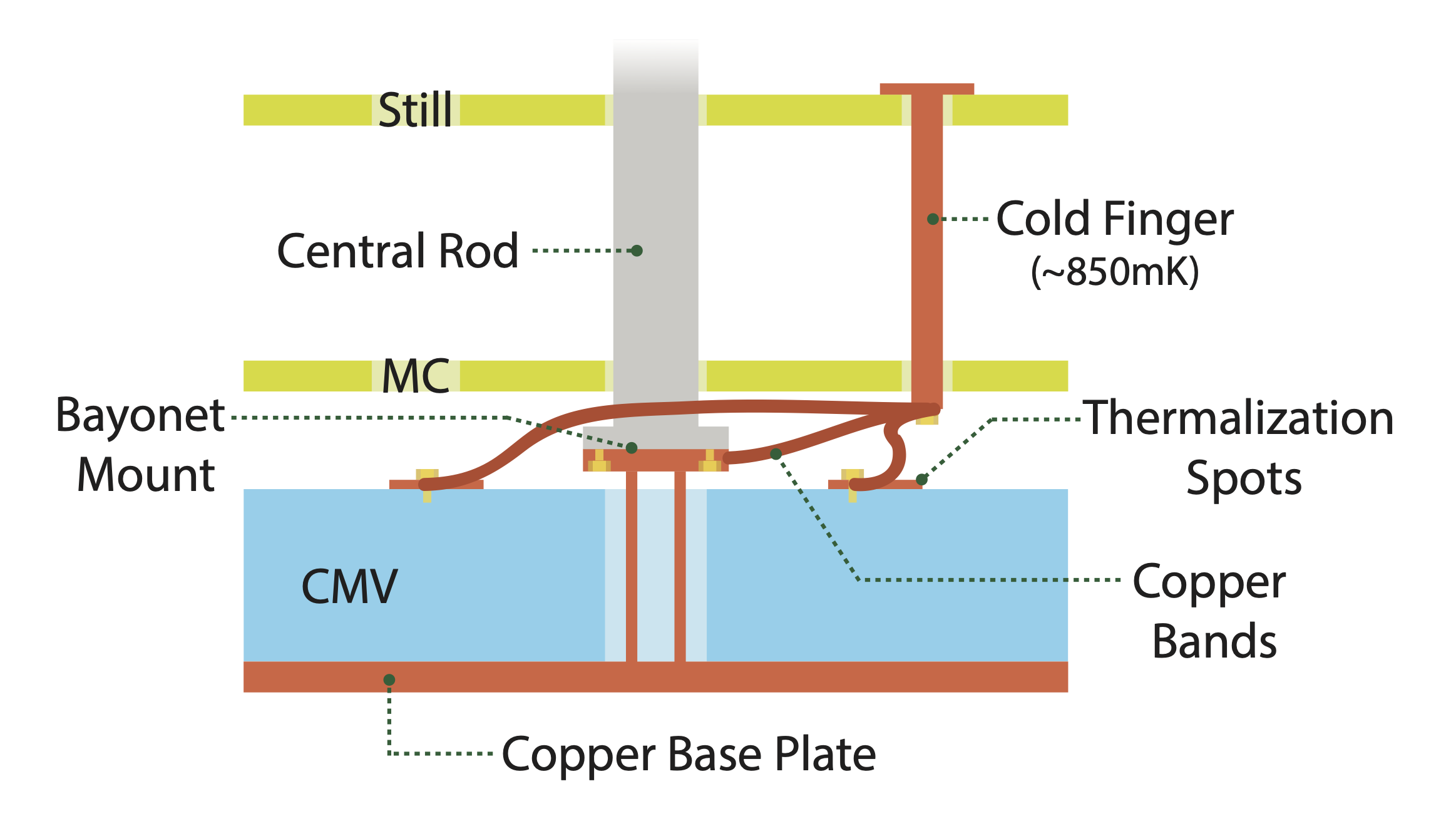}
\caption{2\,D-schematic of the NUCLEUS cryogenic muon veto thermalization concept. The plastic scintillator disc is positioned on a copper base plate, mechanically mounted to the cryostat's central rod using an easy-to-fasten bayonet mount. Several copper thermalization spots, screwed to the plastic, are thermally coupled to a copper cold finger extending below the still stage via flexible copper bands.}
\label{fig_therma}
\end{center}
\end{figure}

To achieve efficient cooling to still temperature within one week, a tailored thermalization concept was developed and evaluated in a prototype cool-down. The thermalization concept is shown schematically in figure \ref{fig_therma}. The concept involves a copper cold finger (cross-sectional area of 256\,mm$^2$) attached beneath the still stage and extending below the mixing chamber stage. Four individual copper thermalization spots are thermally connected to the plastic scintillator, positioned on top of it. By employing multiple individual copper thermalization spots, rather than forcing a large contact area, the disparate thermal contraction coefficients between copper and polystyrene can be accounted for. Thin copper bands (cross-sectional area of 2.5\,mm$^2$) are connecting the thermalization spots to the cold finger. To mitigate the effects of increasing thermal boundary resistance, the thermalization spots are screwed onto the cryogenic muon veto, ensuring high contact pressure. A thin layer of Apiezon grease \cite{Apiezon} is applied between the surfaces to maximize effective contact area and enhance thermal coupling. The cryogenic muon veto is sitting above the target detectors on a copper base plate, mechanically connected via a designated bayonet mount to a central rod passing through the cryostat. 

To verify the thermalization concept developed for the NUCLEUS cryogenic muon veto, we have tested it by cooling a cylindrical $\sim$\,3\,kg sample of the UPS-923A plastic scintillator inside the cryostat. The measured cool-down curve of the plastic scintillator is depicted in figure \ref{fig_cool-down} in section \ref{sec:therma}. In the prototype run, the plastic scintillator reached still temperature within 5.4\,days of cool-down time. Besides the increasing Kapitza resistance towards lower temperatures and the limited contact pressure that can be exerted on an amorphous polymer compared to a metal, it is primarily the drastically reduced cross-sectional thermal coupling area compared to the hypothetical ideal case that is expected to significantly limit the thermal coupling strength. This explains the prolonged thermalization time of several days, as compared to the $\sim$\,2\,hours delay estimated in section \ref{sec:therma} for the case of ideal coupling. By increasing the number of thermalization spots or the cross-sectional area of the linking copper bands, the thermal coupling strength could be further increased, thereby reducing the cool-down time. Nevertheless, the achieved cool-down time of 5.4\,days with the given realisation of the thermal coupling is yet fully compatible with the anticipated overall cool-down time for the entire NUCLEUS experiment. Due to the presence of a cryostat-internal passive shielding consisting of about 13.5\,kg of polyethylene and 39\,kg of lead (see figure \ref{fig_concept}), a cool-down time of 10\,days or more is anticipated for the NUCLEUS setup. This timeframe is comparable to that of other cryogenic experiments with significant passive shielding at low temperatures \cite{abdelhameed2019first,adams2022cuore}.

%%%%%%%%%%%%%%%%%%%%%%%%%%%%%%%%%%%%%%%%%%%
%%%%%%%%%%%%% Section 3.2 %%%%%%%%%%%%%%%%%
	
    \subsection{Results of the Cryogenic Muon Veto Characterization} 
	\label{sec:performance}

We have installed and commissioned the cryogenic muon veto conceived for the NUCLEUS experiment inside the running pulse-tube pre-cooled dry dilution refrigerator at sub-Kelvin temperatures. Thereby, the fiber-equipped plastic scintillator disk (297\,mm in diameter, 75\,mm in height) has been fully characterized in terms of its capability to separate the ambient gamma background from muons and, accordingly, its muon identification power. This is of central interest for the performance of the muon veto. For the measurements discussed in the following, the assembled cryogenic muon veto was mechanically mounted on a copper plate suspended below the mixing chamber stage of the NUCLEUS cryostat (as shown in the image in figure \ref{fig_concept}). The guidance of the fibers through the dry dilution refrigerator, the mounting of the SiPM to the 300\,K stage and the signal processing have been implemented as described in section \ref{sec:int}. The plastic scintillator disk reached a final operating temperature of 860\,mK, which lies within the nominal base temperature range of the still stage of a Bluefors LD\,400 $^3$He\,/\,$^4$He dry dilution refrigerator. The temperature of the plastic was monitored using a calibrated ruthenium oxide temperature sensor directly screwed into the plastic.

 	\begin{figure*}[t!]
    \centering
    \begin{subfigure}{.5\textwidth}
    \centering
      \includegraphics[width=.93\linewidth]{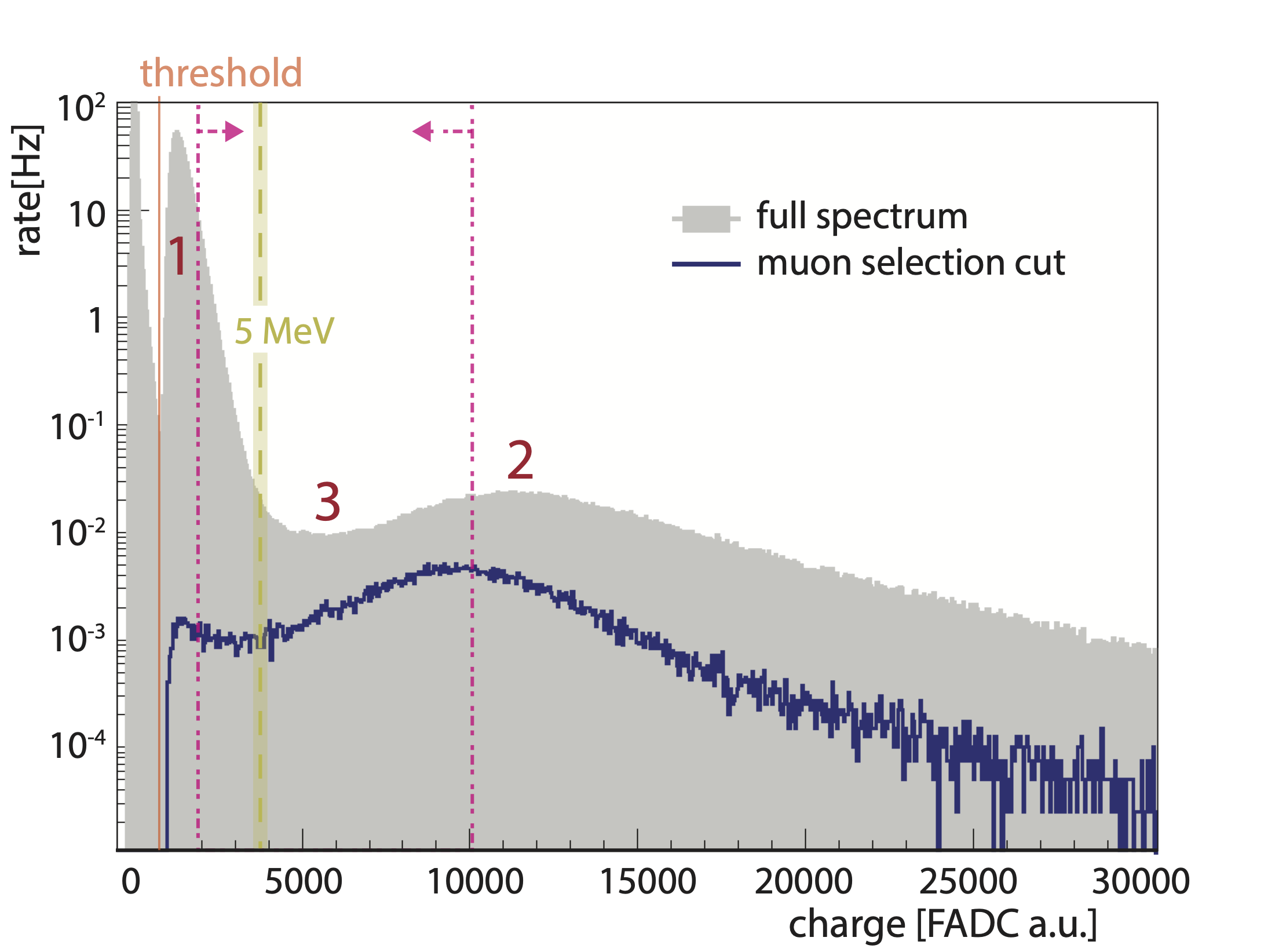}
      \caption{Full charge spectrum and muon cut.}
      \label{fig_spectrum}
    \end{subfigure}%
    \begin{subfigure}{.5\textwidth}
      \centering
      \includegraphics[width=.93\linewidth]{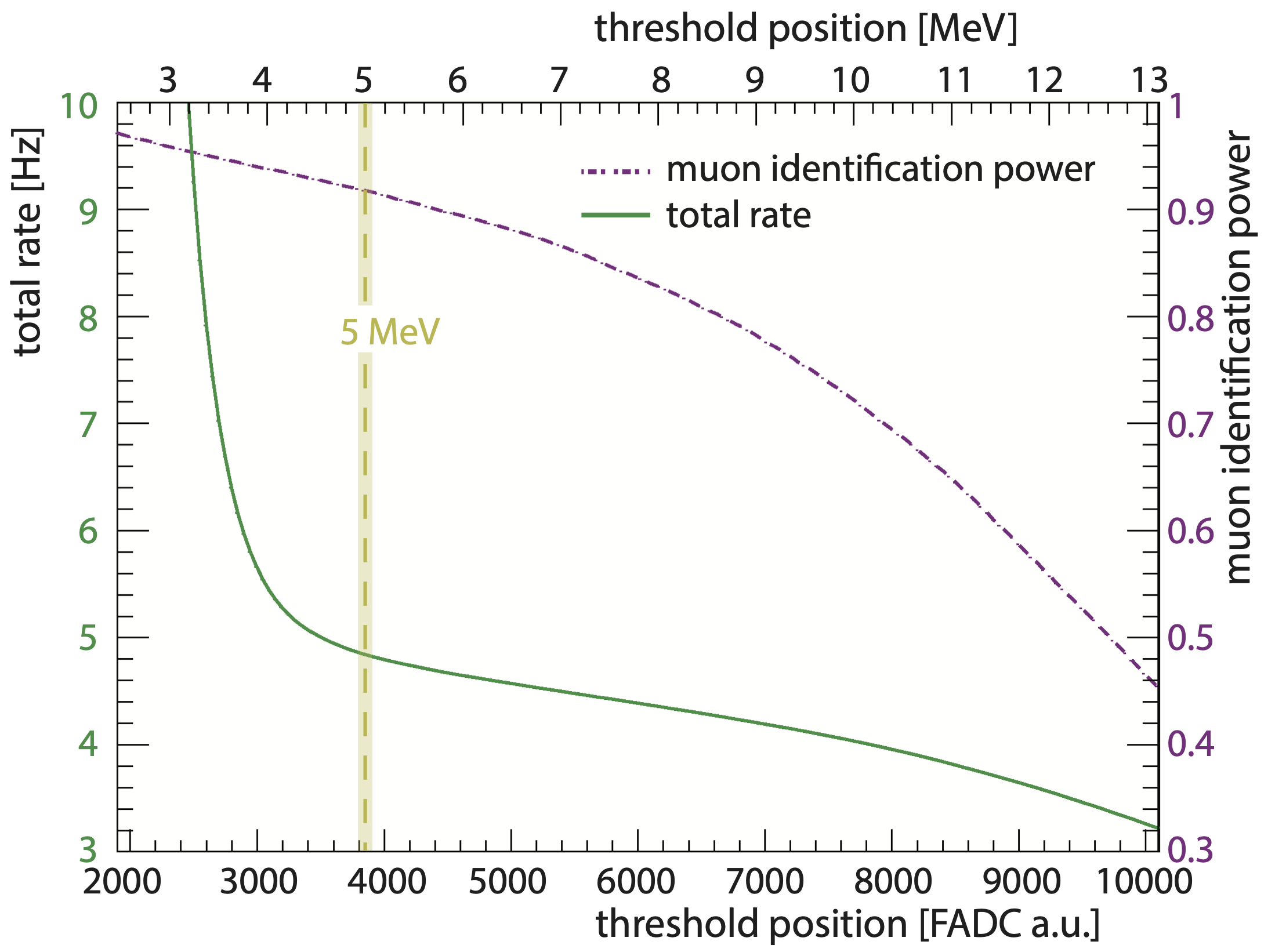}
     \caption{Total rate and muon identification power.}
      \label{fig_MIP}
    \end{subfigure}
    \caption{($\textbf{a}$) shows the full charge spectrum of events (\textit{gray}) recorded with the cryogenic muon veto disk operating at 860\,mK, and a restriction to exclusively muon events via the coincidence criterion (\textit{blue}). The full spectrum features (1) a region for low energetic background events cut at a certain threshold (\textit{orange}), (2) a muon distribution and (3) a well defined “plateau”. The most probable value of the muon distribution can be determined with a Landau fit to be 11550\,$\pm$\,1767\,FADC\,a.u., corresponding to 15\,MeV. A conservative 5\,MeV energy cut is indicated in the plots as a reference ($\textit{yellow}$). ($\textbf{b}$) shows the total rate ($\textit{green}$) and the muon identification power ($\textit{purple}$) as function of the threshold position, retrieved from the corresponding spectra. The inspected region between 2000\,FADC\,a.\,u. and 10000\,FADC\,a.\,u. is marked ($\textit{magenta arrows}$) in the charge spectrum.}
    \label{spectrum_fig}
	\end{figure*}

Figure \ref{fig_spectrum} shows the measured charge spectrum of events in the cryogenic muon veto thermalized at 860\,mK, triggering at a low threshold of a few mV. The full spectrum features (1) an external gamma-background sharply increasing towards the threshold, (2) Landau-like distributed muon events and (3) a well defined "plateau" separating the two contributions. The pedestal (plotted below the threshold) has been obtained from integrating the empty baseline of each captured trace before onset of the triggered signal pulse and is used to monitor the SiPM stability. The muon distribution can be described by a Landau fit with most probable value at (11550\,$\pm$\,1767)\,FADC\,a.u., corresponding to an energy deposition in the plastic scintillator of 15\,MeV. The latter value has been verified in a dedicated GEANT4 Monte Carlo simulation of the energy deposited by atmospheric muons traversing the plastic scintillator disk. Assuming linear energy response, the default 5\,MeV energy cut applied in the GEANT4 Monte Carlo simulations to estimate the residual background level in the NUCLEUS target detector \cite{Goupy_NU22} can be reconstructed from the measured charge in FADC a.u. and is indicated in the plots in figure \ref{spectrum_fig} as benchmark. Muon events are identified by means of a twofold coincidence with an external scintillator panel installed right underneath the outermost vessel outside of the cryostat. The coincidence criterion constrains the angular acceptance of impinging muons and hence their total rate. As expected, restricting the angular acceptance also decreases the mean path traveled by muons through the plastic scintillator. This, in turn, shifts the most probable value of the coincidence-restricted muon distribution towards lower energies, as can be seen in figure \ref{fig_spectrum}. Moreover, since the twofold coincidence condition suppresses the ambient gamma background, the structure of the muon spectrum down to the trigger threshold is revealed. The shape of this spectrum region is determined primarily by muon events traversing the plastic scintillator near an edge, and is thus strongly geometry dependent. Owing to their reduced track length, these clipping muons deposit less energy in the plastic.

The capability of a plastic scintillation muon veto to separate ambient gamma background from muons is essential in order to minimize false veto signals induced by gammas and at the same time maximize the muon identification power, which is defined as the fraction of selected muon events above a certain threshold. For this reason, the choice of the energy cut above which triggered events are considered for the anti-coincidence with the target detector  is of particular relevance for the operational performance of the cryogenic muon veto. Ideally, this cut maximizes the muon identification power, while keeping the total trigger rate (and thus the induced target detector dead time) as low as possible. Figure \ref{fig_MIP} plots the total rate ($\textit{green}$) and the muon identification power ($\textit{purple}$) as function of the threshold position between 2000\,FADC\,a.\,u. and 10000\,FADC\,a.\,u. in the charge spectrum - corresponding to deposited energies between $\sim$\,2.6\,MeV and $\sim$\,13.1\,MeV. The information on the total rate has been extracted from the full charge spectrum of events (figure \ref{fig_spectrum}, $\textit{gray}$), whereas the information on the muon identification power relies on the coincidence-restricted muon spectrum (figure \ref{fig_spectrum}, $\textit{blue}$). Below around 3000\,FADC\,a.u. (corresponding to $\sim$\,4\,MeV), the total rate is increasingly dominated by the high rate of ambient gammas and hence increases steeply with decreasing threshold position. Nevertheless, due to the compactness of the cryogenic muon veto, a muon identification power of $>$\,95\,\% can be achieved at moderate total rate of $<$\,10\,Hz, given the underlying coincidence-restricted sample of muons.  This will allow us to apply an energy cut even below 5\,MeV without significantly increasing the expected overall muon identification rate of $\sim$\,325\,Hz for the full assembly \cite{Wagner_2022}. With such a low threshold, it will be possible to tag many of the clipping muons. Muons that escape detection in the cryogenic muon veto despite the low threshold may eventually be detected in a neighboring panel. 

\vspace{1em}
The optimal choice of the energy cut for the cryogenic muon veto, and from that the overall efficiency of muon identification, will be conclusively determined during the commissioning of the full NUCLEUS muon veto in 2023 under real operating conditions (i.e. for an unconstrained sample of muons and in the interplay of the complete NUCLEUS muon veto).

%%%%%%%%%%%%%%%%%%%%%%%%%%%%%%%%%%%%%%%%%%%
%%%%%%%%%%%%%%% Section 4 %%%%%%%%%%%%%%%%%

	\section{Conclusion and Outlook} 
	\label{sec:conclusion}

Muon-induced events are anticipated to be among the limiting background sources for experiments searching for rare-events at experimental sites located on-surface. To achieve the sensitivity goals, highly efficient and hermetic muon vetos are indispensable. By deploying cryogenic detectors, which typically exhibit slow time response ($\mathcal{O}$(100\,$\mu$s)), many rare-event search experiments face stringent constraints on the tolerable anti-coincident muon rate. Considering the large experimental volume offered by state-of-the-art cryogenic facilities, crucial for achieving temperatures of $\mathcal{O}$(10\,mK), compact muon veto solutions integrated within the cryostat's interior are an appealing option. In this context, we have developed a novel cryogenic muon veto: an instrumented plastic scintillator operating inside a running dry dilution refrigeration. 

For the NUCLEUS CE$\nu$NS experiment, we construct-ed and commissioned a cryogenic muon veto operating at 860\,mK, which achieves a muon identification power of $>$\,95\,\% at moderate total rate of $<$\,10\,Hz. As simulations have shown, the cryogenic muon veto is an essential component for NUCLEUS to tackle the dominant muon-induced backgrounds and, accordingly, towards achieving the required benchmark background index. 

As a foundation for this application, we demonstrate the feasibility of instrumenting and operating organic plastic scintillators for particle detection in sub-Kelvin temperature environments by examining the thermal material and light output characteristics of a plastic scintillation detector. The polystyrene-based UPS-923A exhibits a low thermal conductivity of $k$\,=\,(0.008165\,$\pm$ 0.000133)\,W\,/\,(K$\cdot$m) at 0.8\,K, along with a considerable heat capacity of $C$\,=\,(0.148\,$\pm$\,0.023)\,J\,/\,(K$\cdot$kg) at 1\,K. In comparison, at similar temperatures copper has a four orders of magnitude higher thermal conductivity and a twenty times lower heat capacity. These numbers imply that much more energy needs to be dissipated for the thermalization of amorphous polymers than for the same amount of copper, but with drastically inferior heat transfer. However, due to the amorphous nature and significant thermal contraction of plastic scintillators, it is not primarily the intrinsic material properties but the challenges in implementing an efficient thermal coupling that limit the thermalization time of kg-scale amorphous polymers. Nevertheless, owing to a sophisticated thermalisation concept, we demonstrate the capability of cooling $\mathcal{O}$(kg) of amorphous polymers down to an operating temperature of $\sim$\,850\,mK within a few days. This cooling timeframe aligns with the typical cool-down times for many experiments utilizing cryogenic facilities, thereby enabling the integration of plastic scintillators into the immediate environment of cryogenic detectors. The measured light output properties of the detector, which was found to increase by a factor of 1.61\,$\pm$\,0.05 towards low temperatures, reveal enhanced photon statistics facilitating efficient light read-out. The time response in the sub-Kelvin temperature range is only modestly slowed down by a factor of 1.29\,$\pm$ 0.09, making it significantly faster than cryogenic detectors even at such low temperatures.

We have developed and deployed a novel and versatile detector concept: it is distinguished by the operation of a SiPM at constant room temperature inside a running dry dilution refrigerator; by the flexible guiding of the WLS fibers through the vessels of a cryostat; and by the plastic scintillator operating at sub-Kelvin temperatures. The read-out technique using WLS fibers and a SiPM to guide and detect the scintillation light provides flexibility, vacuum compatibility, and allows for integration into various cryogenic environments. This technology, with its fast ($\mathcal{O}$(100\,ns)) light read-out, may also be an interesting alternative to the conventional bolometric light read-out of scintillating materials using auxiliary ultra-low temperature sensors, particularly when fast timing rather than high energy resolution is of interest. Furthermore, by operating the active detector material in the interior of a dry dilution refrigerator in close proximity of the cryogenic target detector, it becomes possible to achieve compactness and high efficiency alike. The fast time response and easy scalability to large detector masses in the kilogram range even at low temperatures make plastic scintillators particularly well suited for use as anti-coincidence muon veto at sub-Kelvin temperatures. This opens up new perspectives for the search for rare-events with cryogenic detectors at sites lacking substantial overburden.

\section*{Acknowledgments}
%\begin{acknowledgments}
We thank Patrick Champion and Maurice Chapellier for their support and advice during initial prototype tests at DPhN (IRFU, CEA, Universit\'{e} Paris Saclay). Our special thanks go to the skilled technical staff of our laboratories for their excellent work in the production the NUCLEUS cryogenic muon veto and for their contribution to this work. We extend our thanks to Hans Steiger (Johannes Gutenberg Universität Mainz) for his engaging discussions pertaining to this research.

\section*{Funding}
The research was supported by the ERC StG2018-804228 NUCLEUS, by which the NUCLEUS experiment is funded, as well as the SFB1258 ”Neutrinos and Dark Matter in Astro- and Particle Physics”. NUCLEUS members acknowledge additional funding by the DFG through the  Excellence Cluster ORIGINS, by the P2IO LabEx (ANR-10-LABX-0038) in the framework "Investissements d’Avenir" (ANR-11-IDEX-0003-01) managed by the Agence Nationale de la Recherche (ANR), France and by the Austrian Science Fund (FWF) through the "P 34778-N, ELOISE".

\bibliographystyle{apsrev4-2}

\bibliography{apssamp}   % name your BibTeX data base

%apsrev4-2.bst 2019-01-14 (MD) hand-edited version of apsrev4-1.bst
%Control: key (0)
%Control: author (72) initials jnrlst
%Control: editor formatted (1) identically to author
%Control: production of article title (-1) disabled
%Control: page (0) single
%Control: year (1) truncated
%Control: production of eprint (0) enabled
\begin{thebibliography}{41}%
\makeatletter
\providecommand \@ifxundefined [1]{%
 \@ifx{#1\undefined}
}%
\providecommand \@ifnum [1]{%
 \ifnum #1\expandafter \@firstoftwo
 \else \expandafter \@secondoftwo
 \fi
}%
\providecommand \@ifx [1]{%
 \ifx #1\expandafter \@firstoftwo
 \else \expandafter \@secondoftwo
 \fi
}%
\providecommand \natexlab [1]{#1}%
\providecommand \enquote  [1]{``#1''}%
\providecommand \bibnamefont  [1]{#1}%
\providecommand \bibfnamefont [1]{#1}%
\providecommand \citenamefont [1]{#1}%
\providecommand \href@noop [0]{\@secondoftwo}%
\providecommand \href [0]{\begingroup \@sanitize@url \@href}%
\providecommand \@href[1]{\@@startlink{#1}\@@href}%
\providecommand \@@href[1]{\endgroup#1\@@endlink}%
\providecommand \@sanitize@url [0]{\catcode `\\12\catcode `\$12\catcode `\&12\catcode `\#12\catcode `\^12\catcode `\_12\catcode `\%12\relax}%
\providecommand \@@startlink[1]{}%
\providecommand \@@endlink[0]{}%
\providecommand \url  [0]{\begingroup\@sanitize@url \@url }%
\providecommand \@url [1]{\endgroup\@href {#1}{\urlprefix }}%
\providecommand \urlprefix  [0]{URL }%
\providecommand \Eprint [0]{\href }%
\providecommand \doibase [0]{https://doi.org/}%
\providecommand \selectlanguage [0]{\@gobble}%
\providecommand \bibinfo  [0]{\@secondoftwo}%
\providecommand \bibfield  [0]{\@secondoftwo}%
\providecommand \translation [1]{[#1]}%
\providecommand \BibitemOpen [0]{}%
\providecommand \bibitemStop [0]{}%
\providecommand \bibitemNoStop [0]{.\EOS\space}%
\providecommand \EOS [0]{\spacefactor3000\relax}%
\providecommand \BibitemShut  [1]{\csname bibitem#1\endcsname}%
\let\auto@bib@innerbib\@empty
%</preamble>
\bibitem [{\citenamefont {Akimov}\ \emph {et~al.}(2017)\citenamefont {Akimov}, \citenamefont {Albert}, \citenamefont {An}, \citenamefont {Awe}, \citenamefont {Barbeau} \emph {et~al.}}]{akimov2017observation}%
  \BibitemOpen
  \bibfield  {author} {\bibinfo {author} {\bibfnamefont {D.}~\bibnamefont {Akimov}}, \bibinfo {author} {\bibfnamefont {J.}~\bibnamefont {Albert}}, \bibinfo {author} {\bibfnamefont {P.}~\bibnamefont {An}}, \bibinfo {author} {\bibfnamefont {C.}~\bibnamefont {Awe}}, \bibinfo {author} {\bibfnamefont {P.}~\bibnamefont {Barbeau}}, \emph {et~al.},\ }\href {https://doi.org/10.1126/science.aao0990} {\bibfield  {journal} {\bibinfo  {journal} {Science}\ }\textbf {\bibinfo {volume} {357}},\ \bibinfo {pages} {1123} (\bibinfo {year} {2017})}\BibitemShut {NoStop}%
\bibitem [{\citenamefont {Akimov}\ \emph {et~al.}(2021)\citenamefont {Akimov}, \citenamefont {Albert}, \citenamefont {An}, \citenamefont {Awe}, \citenamefont {Barbeau} \emph {et~al.}}]{akimov2021first}%
  \BibitemOpen
  \bibfield  {author} {\bibinfo {author} {\bibfnamefont {D.}~\bibnamefont {Akimov}}, \bibinfo {author} {\bibfnamefont {J.}~\bibnamefont {Albert}}, \bibinfo {author} {\bibfnamefont {P.}~\bibnamefont {An}}, \bibinfo {author} {\bibfnamefont {C.}~\bibnamefont {Awe}}, \bibinfo {author} {\bibfnamefont {P.}~\bibnamefont {Barbeau}}, \emph {et~al.},\ }\href {https://doi.org/https://doi.org/10.1103/PhysRevLett.126.012002} {\bibfield  {journal} {\bibinfo  {journal} {Phys. Rev. Lett.}\ }\textbf {\bibinfo {volume} {126}},\ \bibinfo {pages} {012002} (\bibinfo {year} {2021})}\BibitemShut {NoStop}%
\bibitem [{\citenamefont {Singh}\ and\ \citenamefont {Wong}(2017)}]{Singh:2017jow}%
  \BibitemOpen
  \bibfield  {author} {\bibinfo {author} {\bibfnamefont {L.}~\bibnamefont {Singh}}\ and\ \bibinfo {author} {\bibfnamefont {H.~T.}\ \bibnamefont {Wong}},\ }\href {https://doi.org/10.1088/1742-6596/888/1/012124} {\bibfield  {journal} {\bibinfo  {journal} {J. Phys. Conf. Ser.}\ }\textbf {\bibinfo {volume} {888}},\ \bibinfo {pages} {012124} (\bibinfo {year} {2017})}\BibitemShut {NoStop}%
\bibitem [{\citenamefont {Salagnac}(2020)}]{SalagnacM7:2020}%
  \BibitemOpen
  \bibfield  {author} {\bibinfo {author} {\bibfnamefont {T.}~\bibnamefont {Salagnac}},\ }in\ \href {https://indi.to/63w2C} {\emph {\bibinfo {booktitle} {Magnificent CEvNS workshop}}}\ (\bibinfo {year} {2020})\BibitemShut {NoStop}%
\bibitem [{\citenamefont {Hakenm\"uller}\ \emph {et~al.}(2019)\citenamefont {Hakenm\"uller} \emph {et~al.}}]{Hakenmuller:2019ecb}%
  \BibitemOpen
  \bibfield  {author} {\bibinfo {author} {\bibfnamefont {J.}~\bibnamefont {Hakenm\"uller}} \emph {et~al.},\ }\href {https://doi.org/10.1140/epjc/s10052-019-7160-2} {\bibfield  {journal} {\bibinfo  {journal} {Eur. Phys. J. C}\ }\textbf {\bibinfo {volume} {79}},\ \bibinfo {pages} {699} (\bibinfo {year} {2019})}\BibitemShut {NoStop}%
\bibitem [{\citenamefont {Aguilar-Arevalo}\ \emph {et~al.}(2019)\citenamefont {Aguilar-Arevalo} \emph {et~al.}}]{Aguilar:2019jlr}%
  \BibitemOpen
  \bibfield  {author} {\bibinfo {author} {\bibfnamefont {A.}~\bibnamefont {Aguilar-Arevalo}} \emph {et~al.} (\bibinfo {collaboration} {CONNIE}),\ }\href {https://doi.org/10.1103/PhysRevD.100.092005} {\bibfield  {journal} {\bibinfo  {journal} {Phys. Rev. D}\ }\textbf {\bibinfo {volume} {100}},\ \bibinfo {pages} {092005} (\bibinfo {year} {2019})}\BibitemShut {NoStop}%
\bibitem [{\citenamefont {Agnolet}\ \emph {et~al.}(2017)\citenamefont {Agnolet} \emph {et~al.}}]{Agnolet:2016zir}%
  \BibitemOpen
  \bibfield  {author} {\bibinfo {author} {\bibfnamefont {G.}~\bibnamefont {Agnolet}} \emph {et~al.} (\bibinfo {collaboration} {MINER}),\ }\href {https://doi.org/10.1016/j.nima.2017.02.024} {\bibfield  {journal} {\bibinfo  {journal} {Nucl. Instrum. Meth. A}\ }\textbf {\bibinfo {volume} {853}},\ \bibinfo {pages} {53} (\bibinfo {year} {2017})}\BibitemShut {NoStop}%
\bibitem [{\citenamefont {Akimov}\ \emph {et~al.}(2020)\citenamefont {Akimov} \emph {et~al.}}]{Akimov:2019ogx}%
  \BibitemOpen
  \bibfield  {author} {\bibinfo {author} {\bibfnamefont {D.}~\bibnamefont {Akimov}} \emph {et~al.} (\bibinfo {collaboration} {RED-100}),\ }\href {https://doi.org/10.1088/1748-0221/15/02/P02020} {\bibfield  {journal} {\bibinfo  {journal} {J. Instrum.}\ }\textbf {\bibinfo {volume} {15}}\bibinfo  {number} { (02)},\ \bibinfo {pages} {P02020}}\BibitemShut {NoStop}%
\bibitem [{\citenamefont {Belov}\ \emph {et~al.}(2015)\citenamefont {Belov} \emph {et~al.}}]{Belov:2015ufh}%
  \BibitemOpen
\bibfield  {number} {  }\bibfield  {author} {\bibinfo {author} {\bibfnamefont {V.}~\bibnamefont {Belov}} \emph {et~al.},\ }\href {https://doi.org/10.1088/1748-0221/10/12/P12011} {\bibfield  {journal} {\bibinfo  {journal} {J. Instrum.}\ }\textbf {\bibinfo {volume} {10}}\bibinfo  {number} { (12)},\ \bibinfo {pages} {P12011}}\BibitemShut {NoStop}%
\bibitem [{\citenamefont {Abdelhameed}\ \emph {et~al.}(2019)\citenamefont {Abdelhameed}, \citenamefont {Angloher}, \citenamefont {Bauer} \emph {et~al.}}]{abdelhameed2019first}%
  \BibitemOpen
\bibfield  {number} {  }\bibfield  {author} {\bibinfo {author} {\bibfnamefont {A.~H.}\ \bibnamefont {Abdelhameed}}, \bibinfo {author} {\bibfnamefont {G.}~\bibnamefont {Angloher}}, \bibinfo {author} {\bibfnamefont {P.}~\bibnamefont {Bauer}}, \emph {et~al.},\ }\href {https://doi.org/https://doi.org/10.1103/PhysRevD.100.102002} {\bibfield  {journal} {\bibinfo  {journal} {Physical Review D}\ }\textbf {\bibinfo {volume} {100}},\ \bibinfo {pages} {102002} (\bibinfo {year} {2019})}\BibitemShut {NoStop}%
\bibitem [{\citenamefont {Alkhatib}\ \emph {et~al.}(2021)\citenamefont {Alkhatib}, \citenamefont {Amaral}, \citenamefont {Aralis}, \citenamefont {Aramaki}, \citenamefont {Arnquist} \emph {et~al.}}]{alkhatib2021light}%
  \BibitemOpen
  \bibfield  {author} {\bibinfo {author} {\bibfnamefont {I.}~\bibnamefont {Alkhatib}}, \bibinfo {author} {\bibfnamefont {D.}~\bibnamefont {Amaral}}, \bibinfo {author} {\bibfnamefont {T.}~\bibnamefont {Aralis}}, \bibinfo {author} {\bibfnamefont {T.}~\bibnamefont {Aramaki}}, \bibinfo {author} {\bibfnamefont {I.~J.}\ \bibnamefont {Arnquist}}, \emph {et~al.},\ }\href {https://doi.org/https://doi.org/10.1103/PhysRevLett.127.061801} {\bibfield  {journal} {\bibinfo  {journal} {Phys. Rev. Lett.}\ }\textbf {\bibinfo {volume} {127}},\ \bibinfo {pages} {061801} (\bibinfo {year} {2021})}\BibitemShut {NoStop}%
\bibitem [{\citenamefont {Armengaud}\ \emph {et~al.}(2019)\citenamefont {Armengaud}, \citenamefont {Augier}, \citenamefont {Beno{\^\i}t}, \citenamefont {Benoit}, \citenamefont {Berg{\'e}} \emph {et~al.}}]{armengaud2019searching}%
  \BibitemOpen
  \bibfield  {author} {\bibinfo {author} {\bibfnamefont {E.}~\bibnamefont {Armengaud}}, \bibinfo {author} {\bibfnamefont {C.}~\bibnamefont {Augier}}, \bibinfo {author} {\bibfnamefont {A.}~\bibnamefont {Beno{\^\i}t}}, \bibinfo {author} {\bibfnamefont {A.}~\bibnamefont {Benoit}}, \bibinfo {author} {\bibfnamefont {L.}~\bibnamefont {Berg{\'e}}}, \emph {et~al.},\ }\href {https://doi.org/https://doi.org/10.1103/PhysRevD.99.082003} {\bibfield  {journal} {\bibinfo  {journal} {Phys. Rev. D}\ }\textbf {\bibinfo {volume} {99}},\ \bibinfo {pages} {082003} (\bibinfo {year} {2019})}\BibitemShut {NoStop}%
\bibitem [{\citenamefont {Heusser}(1995)}]{heusser1995low}%
  \BibitemOpen
  \bibfield  {author} {\bibinfo {author} {\bibfnamefont {G.}~\bibnamefont {Heusser}},\ }\href@noop {} {\bibfield  {journal} {\bibinfo  {journal} {Annu. Rev. Nucl. Part. Sci.}\ }\textbf {\bibinfo {volume} {45}},\ \bibinfo {pages} {543} (\bibinfo {year} {1995})}\BibitemShut {NoStop}%
\bibitem [{\citenamefont {Angloher}\ \emph {et~al.}(2019)\citenamefont {Angloher} \emph {et~al.}}]{Angloher:2019flc}%
  \BibitemOpen
  \bibfield  {author} {\bibinfo {author} {\bibfnamefont {G.}~\bibnamefont {Angloher}} \emph {et~al.},\ }\href {https://doi.org/10.1140/epjc/s10052-019-7454-4} {\bibfield  {journal} {\bibinfo  {journal} {Eur. Phys. J. C}\ }\textbf {\bibinfo {volume} {79}},\ \bibinfo {pages} {1018} (\bibinfo {year} {2019})}\BibitemShut {NoStop}%
\bibitem [{\citenamefont {{Bluefors Oy}}(2023)}]{BF_LD}%
  \BibitemOpen
  \bibfield  {author} {\bibinfo {author} {\bibnamefont {{Bluefors Oy}}},\ }\href@noop {} {\bibinfo {title} {{LD 400 Dry Dilution Refrigerator}}},\ \bibinfo {howpublished} {\url{https://bluefors.com/products/ld-dilution-refrigerator/}} (\bibinfo {year} {2023}),\ \bibinfo {note} {[Online; accessed 23-May-2023]}\BibitemShut {NoStop}%
\bibitem [{\citenamefont {{Cryo\,Concept}}(2023)}]{CC_Cryo}%
  \BibitemOpen
  \bibfield  {author} {\bibinfo {author} {\bibnamefont {{Cryo\,Concept}}},\ }\href@noop {} {\bibinfo {title} {{Hexa Dry L Dilution Refrigerator}}},\ \bibinfo {howpublished} {\url{https://cryoconcept.com/product/hexa-dry-l/}} (\bibinfo {year} {2023}),\ \bibinfo {note} {[Online; accessed 23-May-2023]}\BibitemShut {NoStop}%
\bibitem [{\citenamefont {{Oxford Instruments}}(2023)}]{Oxford_Cryo}%
  \BibitemOpen
  \bibfield  {author} {\bibinfo {author} {\bibnamefont {{Oxford Instruments}}},\ }\href@noop {} {\bibinfo {title} {{Proteox\,MX Dilution Refrigerator}}},\ \bibinfo {howpublished} {\url{https://nanoscience.oxinst.com/products/proteoxmx}} (\bibinfo {year} {2023}),\ \bibinfo {note} {[Online; accessed 23-May-2023]}\BibitemShut {NoStop}%
\bibitem [{\citenamefont {Ventura}\ and\ \citenamefont {Risegari}(2010)}]{ventura2010art}%
  \BibitemOpen
  \bibfield  {author} {\bibinfo {author} {\bibfnamefont {G.}~\bibnamefont {Ventura}}\ and\ \bibinfo {author} {\bibfnamefont {L.}~\bibnamefont {Risegari}},\ }\href@noop {} {\emph {\bibinfo {title} {{The art of cryogenics: low-temperature experimental techniques}}}}\ (\bibinfo  {publisher} {Elsevier},\ \bibinfo {year} {2010})\BibitemShut {NoStop}%
\bibitem [{\citenamefont {Bradley}\ \emph {et~al.}(2013)\citenamefont {Bradley}, \citenamefont {Radebaugh} \emph {et~al.}}]{bradley2013properties}%
  \BibitemOpen
  \bibfield  {author} {\bibinfo {author} {\bibfnamefont {P.~E.}\ \bibnamefont {Bradley}}, \bibinfo {author} {\bibfnamefont {R.}~\bibnamefont {Radebaugh}}, \emph {et~al.},\ }\href@noop {} {\bibfield  {journal} {\bibinfo  {journal} {NIST Publ.}\ }\textbf {\bibinfo {volume} {680}},\ \bibinfo {pages} {1} (\bibinfo {year} {2013})}\BibitemShut {NoStop}%
\bibitem [{\citenamefont {Beddar}(2012)}]{beddar2012possible}%
  \BibitemOpen
  \bibfield  {author} {\bibinfo {author} {\bibfnamefont {S.}~\bibnamefont {Beddar}},\ }\href {https://doi.org/http://dx.doi.org/10.1118/1.4748508} {\bibfield  {journal} {\bibinfo  {journal} {Med. Phys.}\ }\textbf {\bibinfo {volume} {39}},\ \bibinfo {pages} {6522} (\bibinfo {year} {2012})}\BibitemShut {NoStop}%
\bibitem [{\citenamefont {Peralta}(2018)}]{peralta2018temperature}%
  \BibitemOpen
  \bibfield  {author} {\bibinfo {author} {\bibfnamefont {L.}~\bibnamefont {Peralta}},\ }\href@noop {} {\bibfield  {journal} {\bibinfo  {journal} {Nucl. Instrum. Meth. A}\ }\textbf {\bibinfo {volume} {883}},\ \bibinfo {pages} {20} (\bibinfo {year} {2018})}\BibitemShut {NoStop}%
\bibitem [{\citenamefont {Buranurak}\ \emph {et~al.}(2013)\citenamefont {Buranurak}, \citenamefont {Andersen} \emph {et~al.}}]{buranurak2013temperature}%
  \BibitemOpen
  \bibfield  {author} {\bibinfo {author} {\bibfnamefont {S.}~\bibnamefont {Buranurak}}, \bibinfo {author} {\bibfnamefont {C.~E.}\ \bibnamefont {Andersen}}, \emph {et~al.},\ }\href {https://doi.org/https://doi.org/10.1016/j.radmeas.2013.01.049} {\bibfield  {journal} {\bibinfo  {journal} {Radiation Measurements}\ }\textbf {\bibinfo {volume} {56}},\ \bibinfo {pages} {307} (\bibinfo {year} {2013})}\BibitemShut {NoStop}%
\bibitem [{\citenamefont {S{\"o}rensen}\ \emph {et~al.}(2018)\citenamefont {S{\"o}rensen}, \citenamefont {Hans}, \citenamefont {Junghans}, \citenamefont {Krosigk}, \citenamefont {K{\"o}gler} \emph {et~al.}}]{sorensen2018temperature}%
  \BibitemOpen
  \bibfield  {author} {\bibinfo {author} {\bibfnamefont {A.}~\bibnamefont {S{\"o}rensen}}, \bibinfo {author} {\bibfnamefont {S.}~\bibnamefont {Hans}}, \bibinfo {author} {\bibfnamefont {A.}~\bibnamefont {Junghans}}, \bibinfo {author} {\bibfnamefont {B.~v.}\ \bibnamefont {Krosigk}}, \bibinfo {author} {\bibfnamefont {T.}~\bibnamefont {K{\"o}gler}}, \emph {et~al.},\ }\href {https://doi.org/https://doi.org/10.1140/epjc/s10052-017-5484-3} {\bibfield  {journal} {\bibinfo  {journal} {Eur. Phys. J. C}\ }\textbf {\bibinfo {volume} {78}},\ \bibinfo {pages} {1} (\bibinfo {year} {2018})}\BibitemShut {NoStop}%
\bibitem [{\citenamefont {{Institute for Scintillation Materials of NAS}}(2023)}]{UPS}%
  \BibitemOpen
  \bibfield  {author} {\bibinfo {author} {\bibnamefont {{Institute for Scintillation Materials of NAS}}},\ }\href@noop {} {\bibinfo {title} {{Plastic Scintillator UPS 923-A}}},\ \bibinfo {howpublished} {\url{http://isma.kharkov.ua/en}} (\bibinfo {year} {2023}),\ \bibinfo {note} {[Online; accessed 23-May-2023]}\BibitemShut {NoStop}%
\bibitem [{\citenamefont {{Apiezon}}(2023)}]{Apiezon}%
  \BibitemOpen
  \bibfield  {author} {\bibinfo {author} {\bibnamefont {{Apiezon}}},\ }\href@noop {} {\bibinfo {title} {{Cryogenic Vacuum N Grease}}},\ \bibinfo {howpublished} {\url{https://apiezon.com/products/vacuum-greases/apiezon-n-grease/}} (\bibinfo {year} {2023}),\ \bibinfo {note} {[Online; accessed 23-May-2023]}\BibitemShut {NoStop}%
\bibitem [{\citenamefont {{National Institute of Standards and Technology }}(2023)}]{mann1977user}%
  \BibitemOpen
  \bibfield  {author} {\bibinfo {author} {\bibnamefont {{National Institute of Standards and Technology }}},\ }\href@noop {} {\bibinfo {title} {{Material Properties: Polystyrene}}},\ \bibinfo {howpublished} {\url{https://trc.nist.gov/cryogenics/materials/Polystyrene/polystyrenerev.html}} (\bibinfo {year} {2023}),\ \bibinfo {note} {[Online; accessed 23-July-2023]}\BibitemShut {NoStop}%
\bibitem [{\citenamefont {Klemens}(1985)}]{klemens1985theory}%
  \BibitemOpen
  \bibfield  {author} {\bibinfo {author} {\bibfnamefont {P.~G.}\ \bibnamefont {Klemens}},\ }\href@noop {} {\bibfield  {journal} {\bibinfo  {journal} {Thermal Conductivity}\ }\textbf {\bibinfo {volume} {18}},\ \bibinfo {pages} {307} (\bibinfo {year} {1985})}\BibitemShut {NoStop}%
\bibitem [{\citenamefont {Pobell}(2007)}]{pobell2007matter}%
  \BibitemOpen
  \bibfield  {author} {\bibinfo {author} {\bibfnamefont {F.}~\bibnamefont {Pobell}},\ }\href@noop {} {\emph {\bibinfo {title} {{Matter and methods at low temperatures}}}}\ (\bibinfo  {publisher} {{Springer Science \& Business Media}},\ \bibinfo {year} {2007})\BibitemShut {NoStop}%
\bibitem [{\citenamefont {{Saint Gobain Crystals}}(2023{\natexlab{a}})}]{SaintGobain2}%
  \BibitemOpen
  \bibfield  {author} {\bibinfo {author} {\bibnamefont {{Saint Gobain Crystals}}},\ }\href@noop {} {\bibinfo {title} {{Wavelength Shifting Fibers BC-91A}}},\ \bibinfo {howpublished} {\url{https://www.crystals.saint-gobain.com/products/scintillating-fiber}} (\bibinfo {year} {2023}{\natexlab{a}}),\ \bibinfo {note} {[Online; accessed 23-May-2023]}\BibitemShut {NoStop}%
\bibitem [{\citenamefont {{Toray Industries}}(2023)}]{TORAY}%
  \BibitemOpen
  \bibfield  {author} {\bibinfo {author} {\bibnamefont {{Toray Industries}}},\ }\href@noop {} {\bibinfo {title} {{Lumirror Foil E6SR}}},\ \bibinfo {howpublished} {\url{https://www.films.toray/en/}} (\bibinfo {year} {2023}),\ \bibinfo {note} {[Online; accessed 23-May-2023]}\BibitemShut {NoStop}%
\bibitem [{\citenamefont {{KETEK}}(2023)}]{KETEK}%
  \BibitemOpen
  \bibfield  {author} {\bibinfo {author} {\bibnamefont {{KETEK}}},\ }\href@noop {} {\bibinfo {title} {{Silicon Photomultiplier PE3325-WB TIA TP}}},\ \bibinfo {howpublished} {\url{https://www.ketek.net/wp-content/uploads/KETEK-SiPM-Module-PE33xx-WB-TIA-xP.pdf}} (\bibinfo {year} {2023}),\ \bibinfo {note} {[Online; accessed 23-May-2023]}\BibitemShut {NoStop}%
\bibitem [{\citenamefont {{Struck}}(2023)}]{Struck}%
  \BibitemOpen
  \bibfield  {author} {\bibinfo {author} {\bibnamefont {{Struck}}},\ }\href@noop {} {\bibinfo {title} {{SIS-3316 FADC,}}},\ \bibinfo {howpublished} {\url{https://www.struck.de/sis3316-2014-03-20.pdf}} (\bibinfo {year} {2023}),\ \bibinfo {note} {[Online; accessed 23-May-2023]}\BibitemShut {NoStop}%
\bibitem [{\citenamefont {Wagner}\ \emph {et~al.}(2022)\citenamefont {Wagner}, \citenamefont {Rogly}, \citenamefont {Erhart} \emph {et~al.}}]{Wagner_2022}%
  \BibitemOpen
  \bibfield  {author} {\bibinfo {author} {\bibfnamefont {V.}~\bibnamefont {Wagner}}, \bibinfo {author} {\bibfnamefont {R.}~\bibnamefont {Rogly}}, \bibinfo {author} {\bibfnamefont {A.}~\bibnamefont {Erhart}}, \emph {et~al.},\ }\href {https://doi.org/10.1088/1748-0221/17/05/t05020} {\bibfield  {journal} {\bibinfo  {journal} {J. Instrum.}\ }\textbf {\bibinfo {volume} {17}}\bibinfo  {number} { (05)},\ \bibinfo {pages} {T05020}}\BibitemShut {NoStop}%
\bibitem [{\citenamefont {Homma}\ \emph {et~al.}(1987)\citenamefont {Homma}, \citenamefont {Murase},\ and\ \citenamefont {Sonehara}}]{homma1987effect}%
  \BibitemOpen
\bibfield  {number} {  }\bibfield  {author} {\bibinfo {author} {\bibfnamefont {Y.}~\bibnamefont {Homma}}, \bibinfo {author} {\bibfnamefont {Y.}~\bibnamefont {Murase}},\ and\ \bibinfo {author} {\bibfnamefont {K.}~\bibnamefont {Sonehara}},\ }\href {https://doi.org/https://doi.org/10.1016/0883-2889(87)90002-5} {\bibfield  {journal} {\bibinfo  {journal} {Int. J. Rad. Appl. Instr. A.}\ }\textbf {\bibinfo {volume} {38}},\ \bibinfo {pages} {91} (\bibinfo {year} {1987})}\BibitemShut {NoStop}%
\bibitem [{\citenamefont {Xia}\ \emph {et~al.}(2014)\citenamefont {Xia}, \citenamefont {Yu}, \citenamefont {Li}, \citenamefont {Sun}, \citenamefont {Ding} \emph {et~al.}}]{xia2014temperature}%
  \BibitemOpen
  \bibfield  {author} {\bibinfo {author} {\bibfnamefont {D.-M.}\ \bibnamefont {Xia}}, \bibinfo {author} {\bibfnamefont {B.-X.}\ \bibnamefont {Yu}}, \bibinfo {author} {\bibfnamefont {X.-B.}\ \bibnamefont {Li}}, \bibinfo {author} {\bibfnamefont {X.-L.}\ \bibnamefont {Sun}}, \bibinfo {author} {\bibfnamefont {Y.-Y.}\ \bibnamefont {Ding}}, \emph {et~al.},\ }\href {https://doi.org/10.1088/1674-1137/38/11/116001} {\bibfield  {journal} {\bibinfo  {journal} {Chin. Phys. C}\ }\textbf {\bibinfo {volume} {38}},\ \bibinfo {pages} {116001} (\bibinfo {year} {2014})}\BibitemShut {NoStop}%
\bibitem [{\citenamefont {Valeur}\ and\ \citenamefont {Berberan-Santos}(2012)}]{valeur2012molecular}%
  \BibitemOpen
  \bibfield  {author} {\bibinfo {author} {\bibfnamefont {B.}~\bibnamefont {Valeur}}\ and\ \bibinfo {author} {\bibfnamefont {M.~N.}\ \bibnamefont {Berberan-Santos}},\ }\href@noop {} {\emph {\bibinfo {title} {Molecular fluorescence: principles and applications}}}\ (\bibinfo  {publisher} {John Wiley \& Sons},\ \bibinfo {year} {2012})\BibitemShut {NoStop}%
\bibitem [{\citenamefont {Agostinelli}\ \emph {et~al.}(2003)\citenamefont {Agostinelli} \emph {et~al.}}]{GEANT4:2002zbu}%
  \BibitemOpen
  \bibfield  {author} {\bibinfo {author} {\bibfnamefont {S.}~\bibnamefont {Agostinelli}} \emph {et~al.} (\bibinfo {collaboration} {GEANT4}),\ }\href {https://doi.org/10.1016/S0168-9002(03)01368-8} {\bibfield  {journal} {\bibinfo  {journal} {Nucl. Instrum. Meth. A}\ }\textbf {\bibinfo {volume} {506}},\ \bibinfo {pages} {250} (\bibinfo {year} {2003})}\BibitemShut {NoStop}%
\bibitem [{\citenamefont {Goupy}(2022)}]{Goupy_NU22}%
  \BibitemOpen
  \bibfield  {author} {\bibinfo {author} {\bibfnamefont {C.}~\bibnamefont {Goupy}},\ }in\ \href {https://doi.org/10.5281/zenodo.6767549} {\emph {\bibinfo {booktitle} {Neutrino 2022}}}\ (\bibinfo {year} {2022})\BibitemShut {NoStop}%
\bibitem [{\citenamefont {{Saint Gobain Crystals}}(2023{\natexlab{b}})}]{cement}%
  \BibitemOpen
  \bibfield  {author} {\bibinfo {author} {\bibnamefont {{Saint Gobain Crystals}}},\ }\href@noop {} {\bibinfo {title} {{Optical Cement BC-600}}},\ \bibinfo {howpublished} {\url{https://www.crystals.saint-gobain.com/products/optical-cement}} (\bibinfo {year} {2023}{\natexlab{b}}),\ \bibinfo {note} {[Online; accessed 23-May-2023]}\BibitemShut {NoStop}%
\bibitem [{\citenamefont {Cs{\'a}thy}\ \emph {et~al.}(2011)\citenamefont {Cs{\'a}thy} \emph {et~al.}}]{csathy2011development}%
  \BibitemOpen
  \bibfield  {author} {\bibinfo {author} {\bibfnamefont {J.~J.}\ \bibnamefont {Cs{\'a}thy}} \emph {et~al.},\ }\href {https://doi.org/https://doi.org/10.1016/j.nima.2011.05.070} {\bibfield  {journal} {\bibinfo  {journal} {Nucl. Instrum. Methods Phys. Res.}\ }\textbf {\bibinfo {volume} {654}},\ \bibinfo {pages} {225} (\bibinfo {year} {2011})}\BibitemShut {NoStop}%
\bibitem [{\citenamefont {Adams}\ \emph {et~al.}(2022)\citenamefont {Adams}, \citenamefont {Alduino}, \citenamefont {Alessandria} \emph {et~al.}}]{adams2022cuore}%
  \BibitemOpen
  \bibfield  {author} {\bibinfo {author} {\bibfnamefont {D.}~\bibnamefont {Adams}}, \bibinfo {author} {\bibfnamefont {C.}~\bibnamefont {Alduino}}, \bibinfo {author} {\bibfnamefont {F.}~\bibnamefont {Alessandria}}, \emph {et~al.},\ }\href {https://doi.org/https://doi.org/10.1016/j.ppnp.2021.103902} {\bibfield  {journal} {\bibinfo  {journal} {Progress in Particle and Nuclear Physics}\ }\textbf {\bibinfo {volume} {122}},\ \bibinfo {pages} {103902} (\bibinfo {year} {2022})}\BibitemShut {NoStop}%
\end{thebibliography}%

\end{document}